\documentclass[journal,10pt,oneside]{IEEEtran}
\usepackage{stfloats}
\usepackage{bbm}
\usepackage{amsfonts}
\usepackage{}
\usepackage{color}
\usepackage{mathrsfs}
\usepackage{algorithm}
\usepackage{algorithmic}

\usepackage{multirow}
\usepackage{graphicx,cite,epsfig,amssymb,amsmath}
\allowdisplaybreaks[4]

\renewenvironment{cases}{\left\{\begin{array}[c]{ll}}{\end{array}\right.}

\newtheorem{lemma}{Lemma}

\graphicspath{{figure/}}  

\begin{document}
	\title{Message-Passing Receiver for OCDM over  Multi-Lag Multi-Doppler Channels }
	
	\author{\IEEEauthorblockN{Yun Liu, Fei Ji, Miaowen Wen, Hua Qing}
		
		
		\thanks{Yun Liu is with the School of Internet Finance and Information Engineering, Guangdong University of Finance,  Guangzhou 510521, China (e-mail: yunliu@gduf.edu.cn). Fei ji and Miaowen Wen are with the School of Electronics and Information Engineering, South China University of Technology, Guangzhou 510640, China (e-mail: feiji@scut.edu.cn, miaowen@scut.edu.cn). Hua Qing is with the School of Software, Zhengzhou University of Light Industry, Zhengzhou 450002, China (e-mail: huaqing@zzuli.edu.cn).}}
	
	
	\maketitle
	\begin{abstract}
		As a new candidate waveform for the next generation wireless communications, orthogonal chirp division multiplexing (OCDM) has attracted growing attention for its ability to achieve full diversity in uncoded transmission, and its robustness to narrow-band interference or impulsive noise. Under high-mobility channels with multiple lags and multiple Doppler-shifts (MLMD), the signal suffers doubly selective (DS) fadings in time and frequency domain, and data symbols modulated on orthogonal chirps are interfered by each other. To address the problem of symbol detection of OCDM over MLMD channel, under the assumption that path attenuation factors, delays, and Doppler shifts of the channel are available, we first derive the closed-form channel matrix in Fresnel domain, and then propose a low-complexity method to  approximate it as a sparse matrix. Based on the approximated Fresnel-domain channel, we propose a message passing (MP) based detector to estimate the transmit symbols iteratively. Finally, under two MLMD channels (an underspread channel for terrestrial vehicular communication, and an  overspread channel for narrow-band underwater acoustic communications), Monte Carlo simulation results and analysis are provided to validate its advantages as a promising detector for OCDM. 
	\end{abstract}
	

	\IEEEpeerreviewmaketitle

	\section{Introduction}
	
	The two fundamental obstacles to achieving high data-rate wireless communications are the multipath effect and the Doppler effect of the channel. The former and the latter lead to frequency selectivity and time selectivity of the channel respectively. As we all know, OFDM has been widely adopted by various commercial systems, such as digital video broadcasting (DVB) \cite{DVB},  wireless local area networks (WLAN) (IEEE 802.11) \cite{WLAN}, the Third-Generation Partnership Project (3GPP) Long-Term Evolution (LTE) mobile telecommunication system \cite{LTE}, and the 5th generation (5G) mobile systems \cite{5G}, etc., due to its excellent ability to combat multipath fading of the quasi-static channel by simply using a single-tap per-subcarrier equalizer. However, OFDM is very sensitive to the Doppler shifts, which would cause inter-carrier interference (ICI) and destroy the orthogonality among subcarriers. The condition for ICI-free transmission is that the duration of OFDM block should be shorter than the coherence time of the channel. Unfortunately, in mobile scenarios, the coherence time of the channel can be very short. The spectral efficiency of the corresponding ICI-free OFDM system would be very limited because the guard interval occupies a considerable proportion of the signal block. In order to achieve high spectrum efficiency over time-varying channels, in practice, the symbol duration of OFDM  is usually larger than the channel coherence time, so ICI is inevitable. In the literature, there have been a lot research activities on eliminating the adverse effect of ICI for OFDM \cite{DS_OFDM_ChnnlEst,DS_OFDM_Eql_Est,DS_OFDM_ICI_Eql,DS_MMSE-OFDM}. 
	
	Recently, orthogonal chirp division multiplexing (OCDM) has been proposed as a novel waveform for high rate wireless communications. It was first theoretically and experimentally demonstrated in fiber-optical communication systems \cite{OCDM_ZhaoJian,OCDM_J_LightWave_2016}. Later, the study on OCDM has been extended to terrestrial radio applications \cite{OCDM_SPL_2017, OCDM_FreqD_Pilot_2022_Ma, OCDM_CFO_2022_Ma, OCDM_PAPR_AI_Ma, OCDM_MIMO_ChnnlEst_2023},  underwater acoustic (UWA) communications \cite{OCDM_UWA_OverResampling_ACCESS2022, OCDM_UWA_IOT_2023, OCDM_UWA_shallowWater_ACCESS2022, DS_UWA_ChnnlEst_CL_2021, OCDM_UWA_ACCESS_2020}, and integrated radar and communication systems \cite{OCDM_Radar_Comp_Trans2022, OCDM_Radar_ChnnlEst_2023, OCDM_Radar_SPL_2022}.
	
	For uncoded transmission, OCDM outperforms OFDM and has similar performance to the single carrier block transmission when the guard interval between consecutive signal blocks is longer than time spread of channel \cite{OCDM_ZhaoJian}. It is worth noting that, since OCDM  disperses the energy of each data symbol throughout the time-frequency plane, it has good resistance to narrowband interference and instantaneous pulse interference \cite{OCDM_Analysis_Xiaoli_Ma}. Hence, OCDM has better bit-error-rate (BER) performance than single carrier block transmission when the cyclic prefix (CP) is insufficient or not inserted (CP-free) \cite{OCDM_ZhaoJian, OCDM_MMSE_CPFree_CL_2020}. Moreover, in terms of system achievable rate, OCDM has been proven to be one of the optimal waveforms for time-selective (TS) or frequency-selective (FS) channel, under the following two assumptions that i) the channel state information (CSI) of the system is only available at the receiver, and ii) sufficient iterative detection with perfect feedback is performed at the receiver \cite{OCDM_optimal_waveform_CL_2018, OCDM_MMSE_PIC_Bomfin_2018}. This also can be explained as that, under TS/FS channels, the OCDM signal complies with the equal gain criterion (EGC), that is data symbols should experience equal gain to maximize performance \cite{OCDM_optimal_waveform_CL_2018, OCDM_MMSE_PIC_Bomfin_2018, DS_Robust_Iter_DS_TWC2021}.
	
	For high-speed mobile applications, communication channels are doubly selective (DS) in time and frequency domain. After passing through DS channels, the orthogonality between chirps in OCDM signal is destroyed. Therefore, the data symbols modulated on the chirps will interfere with each other. To address this problem, some progress has been made in the literature. In \cite{DS_Robust_Iter_DS_TWC2021}, using two unique-words (UW) as the preamble and postamble of several data blocks, the authors propose a frequency-domain interpolation based channel estimation method, which is then used  in a decision feedback receiver based on minimum mean square error with parallel interference cancellation (MMSE-PIC).  It has been demonstrated in the simulation that the scheme proposed in \cite{DS_Robust_Iter_DS_TWC2021} can achieve robust performance when the normalized Doppler-spread is less than 0.1275. In \cite{DS_pilot_WCL_YiyinWang_2023}, to capture the channel variation with larger Doppler-spread, using a basis expansion model (BEM) model to characterize the DS channel,  the authors propose a pilot chirp assisted channel estimator base on MMSE criterion. However, the computation complexity of matrix inversion required by MMSE  is $\mathcal{O}(N^3)$, which is usually computationally prohibitive when the block size is large. In \cite{DS_ChannlEst_DD_Domain}, by modeling the DS channel as multiple paths with distinct attenuation factors, delays, and  Doppler-shifts, the authors propose a low-complexity channel estimation method, in which the pilot is inserted in Fresnel-domain data symbols and the estimation is calculated in delay-Doppler domain.
	
	
	In this paper, similar to \cite{DS_ChannlEst_DD_Domain}, we also focus on OCDM over DS channel with multiple-lags and multiple Doppler-shifts (MLMD). The MLMD model is suitable for a variety of scenarios, such as high-speed train communications,  vehicle-to-infrastructure and vehicle-to-vehicle communications, or narrowband underwater acoustic communications for unmanned undersea vehicle (UUV) \cite{MLMD_highspeedTrain, MLMD_Vehicle, MLMD_UWA}. To demodulate the received signals under MLMD channels, a widely used method is to equalize the distorted signal in the frequency domain by an MMSE equalizer and then convert it to the Fresnel domain for symbol detection. Different from traditional frequency domain equalization based receiver, here we proposed a message passing (MP) based iterative receiver working in the Fresnel domain.
	
	The main contributions of this article are as follows:
	
	\begin{itemize}
		\item given the path gains, delays, and doppler shifts of the baseband equivalent channel,  we derive the closed-form expression of the Fresnel domain channel matrix of the OCDM system, and propose a low-complexity method to approximate it as a sparse matrix, which can be exploited by the receiver to reduce the complexity of symbol detection. 
		
		\item using the approximated Fresnel-domain channel matrix, we describe the input-output relation of OCDM by a factor graph in Fresnel-domain, and then propose a message passing based receiver to detect the data symbols iteratively. 
		
		\item we assess the BER performance of the proposed MP receiver by comparing it with that of the one using MMSE frequency-domain equalizer, under both under-spread (i.e., the coherence time is much larger than the delay spread) and over-spread (i.e. the coherence time is comparable or even shorter than the delay spread) MLMD channels.  
	\end{itemize}

	The remainder of this paper is organized as follows. We first describe the system model of OCDM under MLMD channels in Section~II. Then, we derive the input-output relation of OCDM in the Fresnel domain in Section~III.  We propose a message passing algorithm for OCDM symbol detection in Section~IV. In Section~V, we demonstrates the simulation results. Finally, the paper is concluded in Section VI.
	
	\textit{Notation}: Bold upright letters in uppercase are used to denote matrices, for instance, $\mathbf{A}$, while Bold italic letters in lowercase denote vectors, for instance, $\boldsymbol{a}$. $x(\cdot)$ and $x[\cdot]$ stand for functions with continuous and discrete variable, respectively.  Some mathematical notations are listed as follows.
	\begin{table}[hp] \normalsize 
		\begin{tabular}{ll} 
			$j$ & $\sqrt{-1}$ \\
			$ p(\cdot) $  &  probability density function of an event  \\
			$ \rm{Pr}(\cdot)$ &  probability of an event\\
			$\mathbb{E}(\cdot)$ & expectation of a random variable \\
			$ (\cdot)^{H} $ & Hermitian transpose of a matrix\\
			$ (\cdot)^{T} $ & transpose of a matrix \\
			$ (\cdot)^{-1} $ & inversion of a matrix \\
			$ (\cdot)^{*} $ & conjugate of a complex variable \\
			$\mathrm{diag}(\cdot) $ & diagonal matrix converted from a vector\\
			$\mathcal{Z} $ & the set of integer \\
			$\mathcal{R} $ & the set of real numbers \\
			$\mathbf{I}_n$ & the $ N\times N$ identity matrix \\
			$\left[ \mathbf{A} \right] _{m,n}$ & the $(m,n)$-th element of matrix $\mathbf{A}$ \\
			$\left[ \boldsymbol{a} \right] _{m}$ & the $m$-th element of vector $\boldsymbol{a}$ \\
			$\Re(\cdot)$ & the real part of a complex number \\
			$\Im(\cdot)$ & the imaginary part of a complex number \\
			$\delta \left( \cdot \right)$ & the Dirac delta function \\
			$ \lfloor \cdot \rfloor$ & the largest integer not greater the given number\\
			$\left( \left( n \right) \right) _N$ & $n$ modulo $N$
			
		\end{tabular}
	\end{table}

	\section{System Model}
	\subsection{OCDM Modulation}
\begin{figure}[!t]
		\includegraphics[width=8.8cm]{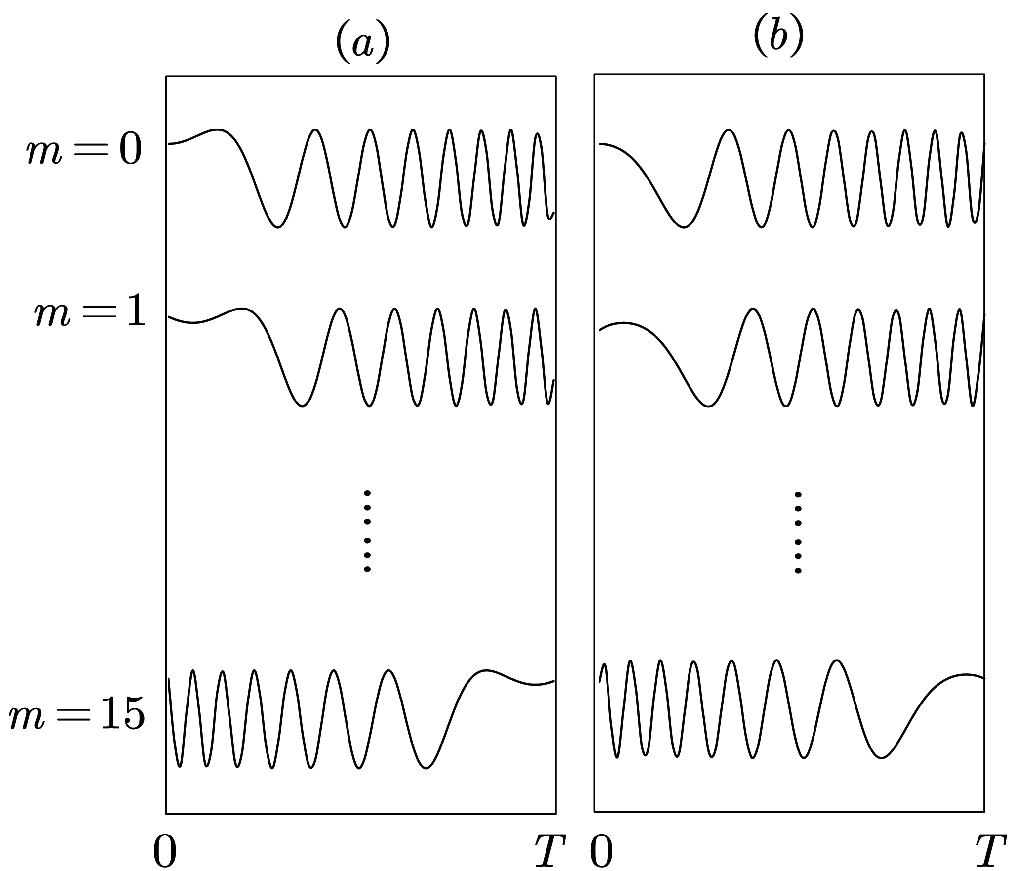}
		\caption{\small Waveforms of a set of orthogonal chirps defined in (\ref{eqn_def_chirp}) when $N=16$. (a) the real part of $\psi _m\left( t \right)$, (b) the imaginary part of $\psi _m\left( t \right)$.}
		\label{figure:chirps-waveform}
	\end{figure}

	Consider an OCDM system of $N$ chirps, and $N$ is an integer power of two. 
	At the transmitter, the incoming information bits are first mapped to independent $M$-ary constellation symbols, which are then split into blocks of length $N$. After that, for each block, all the constellation symbols are modulated onto a parallel of orthogonal chirps, which are superimposed in the time domain (TD). The $m$th, $m$ =  0, 1, $\cdots$, $N-1$, chirp waveform is defined as
	\begin{equation}\label{eqn_def_chirp}
		\psi _m\left( t \right) =e^{j\frac{\pi}{4}}e^{-j\pi \frac{N}{T^2}\left( t-m\frac{T}{N} \right) ^2}, 0\leqslant t<T
	\end{equation}
	where $T$ is the length of the OCDM symbol. All those  chirps are orthogonal to each other. Namely, given any two chirps, $\psi _{m_1}\left( t \right)$ and $\psi _{m_2}\left( t \right)$, we have
	\begin{equation}\label{eqn_chirp_feature}
		\int_0^T{\psi _{m_1}^{*}\left( t \right)}\psi _{m_2}\left( t \right) dt=\begin{cases}
			T, m_1=m_2\\
			0, m_1\ne m_2\\
		\end{cases}.
	\end{equation}
	Fig. \ref{figure:chirps-waveform} presents the waveforms of a family of orthogonal chirps defined in (\ref{eqn_def_chirp}) when $N=16$. The real part and the imaginary part of the complex-valued chirps are given in Fig. \ref{figure:chirps-waveform} (a) and  Fig. \ref{figure:chirps-waveform} (b), respectively.

	\begin{figure*}[!ht]
		\includegraphics[width=18cm]{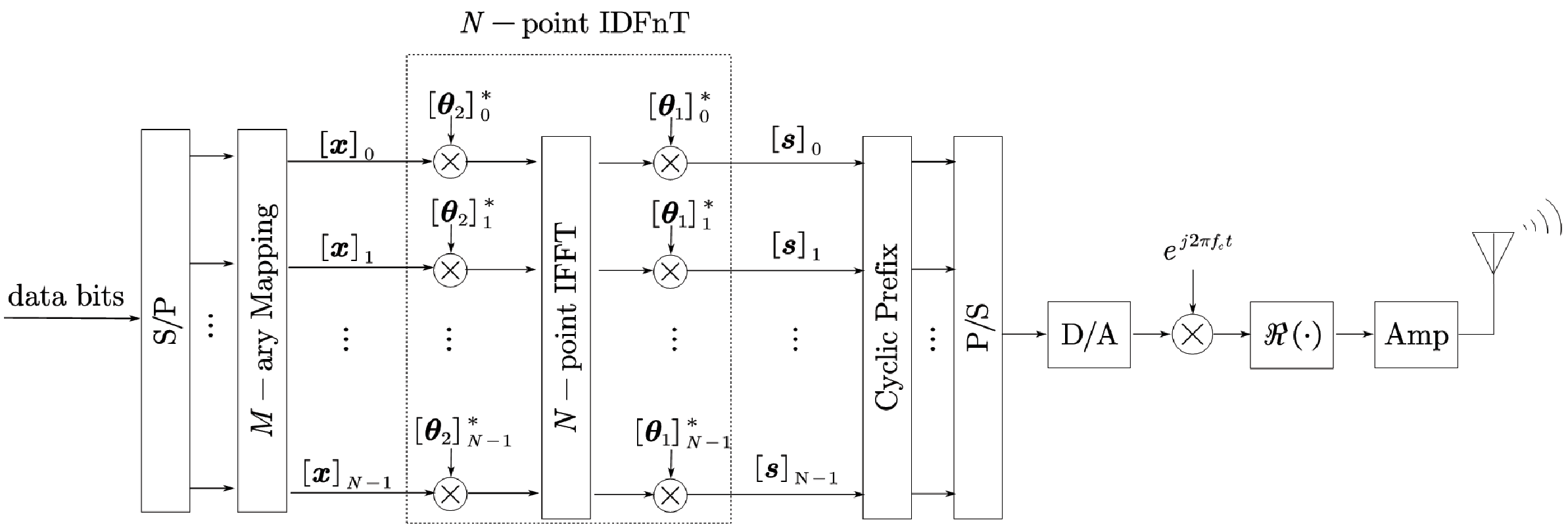}
		\caption{\small The block diagram of OCDM transmitter.}
		\label{diagram_OCDM}
	\end{figure*}
	
	Since the OCDM signals are generated and detected block-by-block, we take a single block as an example to introduce the system, and omit the block index here. Let $x\left[ m \right]$ denote the constellation symbol modulated on the $m$th chirp, $\psi _m\left( t \right) $. The constellation set is defined as $\mathcal{X} =\left\{ \alpha _0, a_1,\cdots ,\alpha _{M-1} \right\} $. The continuous complex-baseband signal of OCDM can be described as
	\begin{equation}\label{eqn_txSig_baseband}
		s\left( t \right) =\,\,\sum_{m=0}^{N-1}{x\left[m \right]  \psi _m\left( t \right) ,}\,\,0\leqslant t<T
	\end{equation}
	where $x\left[ m \right] \in \mathcal{X}$.
	
	The above continuous waveform of OCDM is usually generated from its discrete complex-baseband samples by means of digital-to-analog (D/A) conversion. Corresponding to the chirp $\psi _m\left( t \right)$ defined in (\ref{eqn_def_chirp}), the $m$-th discrete chirp can be written as
	\begin{align}\label{eqn_discrete_chirp-2}
		\psi _m\left[ n \right] &= 
		\sqrt{\frac{T}{N}}\left. \psi _m\left( t \right) \right|_{t=\frac{nT}{N}} \nonumber \\
		& =  \sqrt{\frac{T}{N}}e^{j\frac{\pi}{4}}e^{-j\frac{\pi}{N}\left( n-m \right) ^2}\text{,}n=0,1,\cdots ,N-1.
	\end{align}
	where $T/N$ is the sampling interval. It should be pointed out that, in this paper,  when we convert a continuous-time signal to the corresponding discrete-time one,  we multiply the samples by a constant attenuation coefficient $\sqrt{\frac{T}{N}}$ to keep them having the same signal energy. For instance, as to the signals $\psi _m\left( t \right)$ and $\psi _m\left[ n \right] 
	$ shown in ($\ref{eqn_discrete_chirp-2}$), we have $\int_0^T{\left| \psi _m\left( t \right) \right|^2dt=\sum_{n=0}^{N-1}{\left| \psi _m\left[ n \right] \right|^2}}$. And it is easy to verify that the analog-to-digital (A/D) conversion keeps the orthogonality of the chirps, namely,
	\begin{equation}\label{eqn_discrete_orthorgnality}
		\sum_{n=0}^{N-1}{\psi _{m_1}^{*}\left[ n \right] \psi _{m_2}\left[ n \right] =\begin{cases}
				T,m_1=m_2\\
				0,m_1\ne m_2\\
		\end{cases}}.
	\end{equation}
	
	From  (\ref{eqn_txSig_baseband}), the discrete-time complex-baseband signal of OCDM can be described as 
	\begin{align}\label{eqn_sig_time}
		s\left[ n \right] &= \left. \sqrt{\frac{T}{N}}s\left( t \right) \right|_{t=\frac{nT}{N}} \\
		&= \sum_{m=0}^{N-1}{x\left[ m \right] \psi _m\left[ n \right] }, \text{,} n=0,1,\cdots ,N-1.	
	\end{align}
	
	Let us define vectors $ \boldsymbol{x}=\left[ x\left[ 0 \right] ,x\left[ 1 \right] ,\cdots ,x\left[ N-1 \right] \right] ^T $  and   $
	\boldsymbol{s}\,\,=\left[ s\left[ 0 \right] ,s\left[ 1 \right] ,\cdots ,s\left[ N-1 \right] \right] ^T
	$
	to  denote the transmit discrete signal in the Fresnel domain and the time domain, respectively. Then we have
	\begin{equation}\label{eqn_SX}
		\boldsymbol{s}=\mathbf{\Phi }^H \boldsymbol{x},
	\end{equation}
	where $\mathbf{\Phi}$ is an $N\times N$ discrete Fresnel transform (DFnT) matrix, with the $(m,n)$-th element being $\left[ \mathbf{\Phi } \right] _{m,n}={\color[RGB]{240, 240,240} \frac{1}{\sqrt{T}}}\varphi _{m}^{*}\left[ n \right]$, and $\mathbf{\Phi }^H$
	is  the corresponding inverse discrete Fresnel transform (IDFnT) matrix with $\mathbf{\Phi }^H\mathbf{\Phi }=\mathbf{I}_N$.

	It is worth noting that DFnT can be implemented with a low computational complexity of the order of $N\log _2N
	$  with the help of fast Fourier transform (FFT). Specifically, we can decompose the DFnT matrix as
	\begin{equation} \label{eqn:Fi_LowComplexity}
		\mathbf{\Phi }=\mathbf{\Theta }_2\mathbf{F\Theta }_1,
	\end{equation}
	where 
	$\mathbf{\Theta }_1=\mathrm{diag}\left( \boldsymbol{\theta }_1 \right) $
	and $\mathbf{\Theta }_2=\mathrm{diag}\left( \boldsymbol{\theta }_2\right) $ are diagonal matrices generated by vectors $\boldsymbol{\theta}_1$ and $\boldsymbol{\theta}_2$, with the $m$th element being
	\begin{equation}
		\left[ \boldsymbol{\theta}_1 \right] _{m}=\,\,e^{-j\frac{\pi}{4}}e^{j\frac{\pi}{N}m^2},
	\end{equation}
	and
	\begin{equation}
		\left[ \boldsymbol{\theta}_2 \right] _{m}=e^{j\frac{\pi}{N}m^2},
	\end{equation}
	respectively, and $\mathbf{F}$ is the normalized $N$-point discrete Fourier transform (DFT) matrix with elements $
	\left[ \mathbf{F} \right] _{m,n}=\frac{1}{\sqrt{N}}e^{-j2\pi mn/N}$. It is well known that the number of complex multiplications of $N$-point DFT can be reduced from $N^2$ to $N\log _2N$ by using the FFT algorithm. Similarly, since the IDFnT matrix can be written as $\mathbf{\Phi }^H=\mathbf{\Theta }_{1}^{H}\mathbf{F}^H\mathbf{\Theta }_{2}^{H}$, the computational complexity of IDFnT can also be greatly reduced by using IFFT.

	The bock diagram of complexity-reduced  OCDM transmitter is shown in Fig. \ref{diagram_OCDM}. At the serial to parallel (S/P) converter, the stream of input data bits is split into groups, each containing $N\log _2M$ bits. Then those bits groups are proceeded one by one. For each group of data bits, at first, they are Gray mapped into $N$ symbols, each of which is drawn from a $M$-ary constellation, for instance, quadrature amplitude modulation (QAM) or phase-shift keying (PSK) constellation.  Those symbols are denoted as Fresnel domain vector $\boldsymbol{x}$, which is then transformed into time domain vector $\boldsymbol{s}$, using the complexity reduced IDFnT presented in  (\ref{eqn:Fi_LowComplexity}). To avoid inter-symbol-interference (ISI) induced by the time-spreading of the channel,  a cyclic prefix (CP) of length $T_g$, larger than the time-spread of the channel, is appended to the beginning of $\boldsymbol{s}$. After that, through a  parallel to serial (P/S) converter, the vector is converted into a sequential signal,  which is then changed to a continuous time signal by a analog-to-digital (A/D) converter. So far, the equivalent complex baseband signal of an OCDM symbol is generated. Then it is transformed into a complex bandpass signal by multiplying it with a carrier $e^{j2\pi f_ct}$, and then converted to a real bandpass signal. At last, after amplified by a power amplifier, the continuous time bandpass OCDM signal is sent to the channel by a radio antenna  or an acoustic transducer. The transmit signal can be written as
	\begin{equation}
		\tilde{s}_c\left( t \right) =\mathrm{Re}\left\{ \tilde{s}\left( t \right) e^{j2\pi f_ct} \right\} ,t\in \left[ -T_g,T \right], 
	\end{equation}
	where $\tilde{s}\left( t \right)$ is the complex baseband signal with CP, described as
	\begin{equation}
		\tilde{s}\left( t \right) =\begin{cases}
			s\left( t \right) , t\in \left[ 0, T\right]\\
			s\left( t+T \right) ,t\in \left[ -T_g,0 \right]\\
		\end{cases}.
	\end{equation}

	\subsection{Channel Model}
	\begin{figure}[!ht]
		\includegraphics[width=8cm]{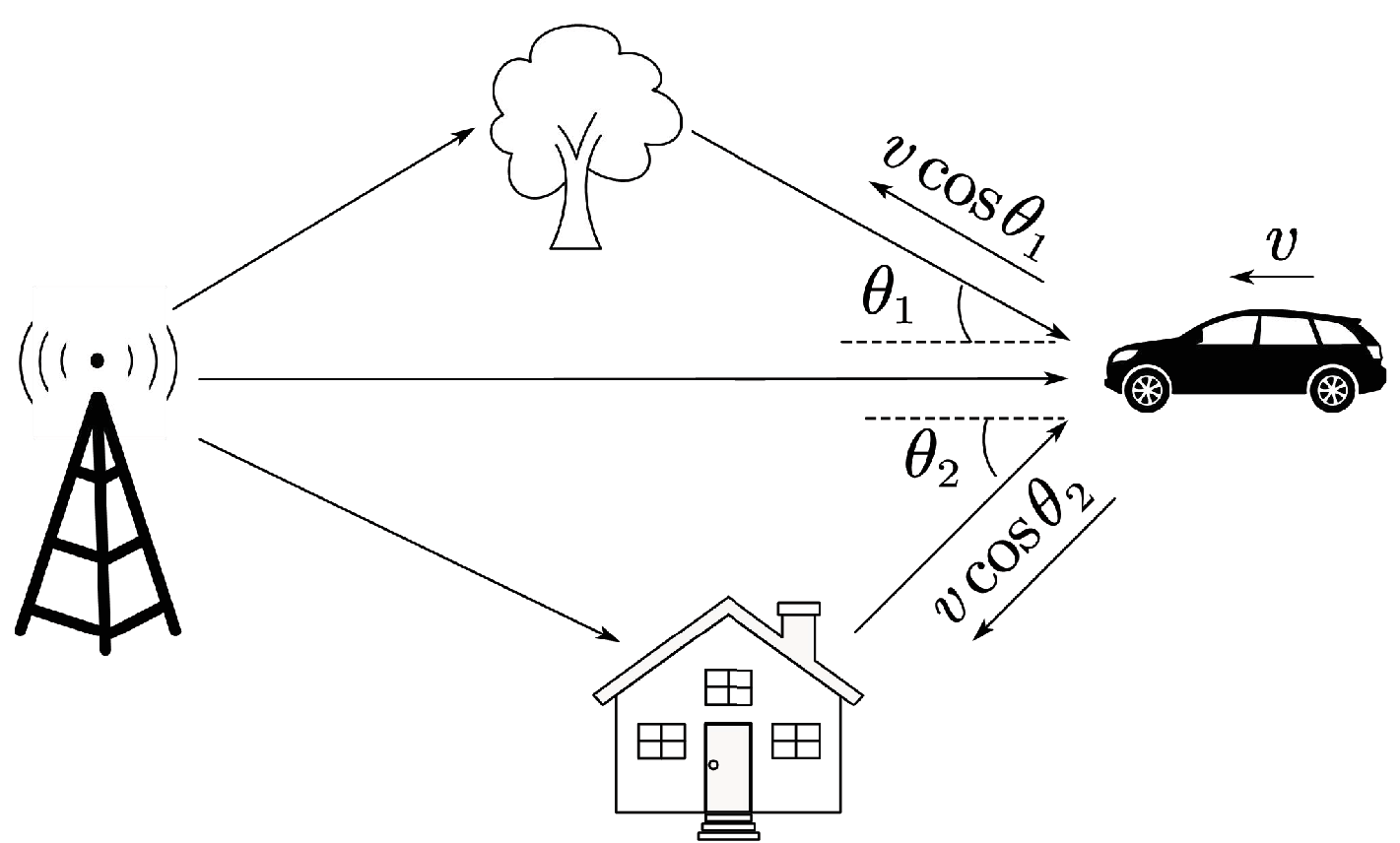}
		\caption{\small Linear time variant channel with multiple lags and multiple Doppler-shifts.}
		\label{fig:mlmd_chnnl}
	\end{figure}
	
	In many wireless communication systems, due to the relative motion between the transmitter and the receiver, the channel can vary significantly even within the time duration of a symbol block. In this case, the channel should be modeled as a linear time-variant (LTV) system. In order to better describe the mathematical model of this kind of LTV channels, as an example, we consider the scenario illustrated in Fig. \ref{fig:mlmd_chnnl}, where the  receiver moves towards the fixed transmitter with speed $v$. There are a number of reflectors in the propagation environment. Using the technique of ray-tracing, we can model the signal arriving at the receiver as the superposition of transmit signal copies from different paths, each characterized by a  delay, a Doppler shift, and an attenuation coefficient \cite{LTVChnnlBook}.  The Doppler shift of the $i$-th path is 
	\begin{equation}
		v_i = f_cV\cos \theta _i/C,
	\end{equation}
	where $C$ is the propagation speed of the wireless medium; $f_c$ is the carrier frequency of the system; and $\theta _i$ is the angle between the incident ray of the $i$-th path and the motion direction of the receiver. Let us denote the delay and attenuation coefficient (corresponding to the complex baseband signal) of the $i$th path as $\tau_i$ and $h_i$. Then, the equivalent complex baseband  channel impulse response (CIR) can be expressed as 
	\begin{equation} \label{eqn:CIR}
		h\left( \tau ,v \right) \,\,=\,\,\sum_{i=1}^P{h_i\delta \left( \tau -\tau _i \right) \delta \left( v-v_i \right)}.
	\end{equation}
	where $P$ is the number of paths.
	Based on (\ref{eqn:CIR}), the received complex bandpass  signal can be written as
	\begin{align} \label{eqn:rttilt}
		r\left( t \right) & =  \iint{h\left( \tau ,v \right) \tilde{s}\left( t-\tau \right) e^{j2\pi v\left( t-\tau \right)}d\tau dv}+\omega\left( t \right)  \nonumber \\
		& = \sum_{i=1}^P{h_i\tilde{s}\left( t-\tau _i \right)}e^{j2\pi v_i\left( t-\tau _i \right)}+\omega\left( t \right) , t\in \left[ 0,T \right] , 
	\end{align}
	where $\omega\left( t \right)$ is complex baseband additive white Gaussian noise (AWGN), independent of the transmit signal and the CIR, with zero mean and power spectral density $N_0$.  
	
	Let us denote the ideal sampling intervals in the time domain and the frequency domain as $T_s = T/N$ and $\Delta f= 1/T$, respectively. Considering the resolution of the sampling interval $T_s$ is generally high enough to  approximate a path delay to its nearest sampling point in typical  communication systems \cite{TseBook}, we write the delay of the $i$th path as
	\begin{equation} \label{eqn:tau}
		\tau _i=\frac{l_iT}{N},
	\end{equation}
	where $l_i$ is an integer, and can be explained as the path delay in discrete time domain. For the Doppler shift $v_i$, we express it as the summation of an integer multiples of frequency interval $\Delta f$ and a fractional one, namely,
	\begin{equation} \label{eqn:Doppler}
		v_i=\frac{\left( k_i+\kappa _i \right)}{T}
	\end{equation}
	where $k_i\in \mathcal{Z}$ and $\kappa _i\in \left(-0.5, 0.5 \right]$. Now, if we substitute (\ref{eqn:tau}) and (\ref{eqn:Doppler}) into (\ref{eqn:rttilt}), the received complex baseband signal can be rewritten as
	\begin{equation}
		r\left( t \right) =\sum_{i=1}^P{h_i\tilde{s}\left( t-\frac{T}{N}l_i \right) e^{j2\pi \frac{k_i+\kappa _i}{T}\left( t-\frac{T}{N}l_i \right)}} + \omega\left( t \right).
	\end{equation}
	
	Under the assumption of ideal timing synchronization at the receiver, with the CP removed, the received discrete complex baseband signal, namely, samples of $r \left(t\right)$, is given by
	\begin{align} \label{eqn:r_n}
		r\left[ n \right] &= \sqrt{\frac{T}{N}}  \left. r\left( t \right) \right|_{t=\frac{nT}{N}} \nonumber \\
		=&\sqrt{\frac{T}{N}}\sum_{i=1}^P{h_i\tilde{s}\left( \frac{T}{N}n-\frac{T}{N}l_i \right) e^{j2\pi \frac{k_i+\kappa _i}{T}\left( \frac{T}{N}n-\frac{T}{N}l_i \right)}} \nonumber \\
		& +\sqrt{\frac{T}{N}}\omega \left( \frac{nT}{N} \right) 
		\nonumber \\
		=& \sum_{i=1}^P{\tilde{h}_ie^{j\frac{2\pi}{N}\left( k_i+\kappa _i \right) n}\tilde{s}\left[ n-l_i \right]} +\omega \left[ n \right]   \nonumber \\
		=& \sum_{i=1}^P{\tilde{h}_ie^{j\frac{2\pi}{N}\left( k_i+\kappa _i \right) n}s\left[ \left( \left( n-l_i \right) \right) _N \right]}  +\omega \left[ n \right] , \nonumber \\
		& n=0,1,\cdots ,N-1, 
	\end{align}
	where $\tilde{h}_i$ is the equivalent path gain of the $i$-th path,  
	\begin{equation}
		\tilde{h}_i=h_ie^{-j\frac{2\pi}{N}\left( k_i+\kappa _i \right) l_i},
	\end{equation}
	$\tilde{s}\left[ n \right]$ is the discrete baseband transmit signal with CP, 
	\begin{equation}
		\tilde{s}\left[ n \right] \,\,=\,\,\begin{cases}
			s\left[ n \right] ,n=0, 1,\cdots N-1\\
			s\left[ n+N \right] , \lfloor \frac{T_g}{T}N \rfloor <n<0\\
		\end{cases},
	\end{equation}  
	and the complex AWGN terms $\omega\left[n\right]$ are independent identically distributed (IID) with with zero mean and variance $N_0$.
	
	Define the received discrete baseband signal and the additive noise contained in it as vectors
	\begin{equation}
		\boldsymbol{r}=\left[ r\left[ 0 \right] ,r\left[ 1 \right] ,\cdots ,r\left[ N-1 \right] \right] ^T,
	\end{equation}
	and 
	\begin{equation}
		\boldsymbol{\omega}=\left[ \omega \left[ 0 \right] ,\omega \left[ 1 \right] ,\cdots ,\omega \left[ N-1 \right] \right] ^T,
	\end{equation}respectively.
	Then, according to (\ref{eqn:r_n}), the channel input-output relationship in the time domain can be written in matrix-vector representation as
	\begin{equation} \label{eqn:v_rcvsig}
		\boldsymbol{r}=\mathbf{H}\boldsymbol{s} + \boldsymbol{\omega}
	\end{equation}
	where 
	\begin{equation} \label{eqn:v_chnnl}
		\mathbf{H}=\sum_{i=1}^P{\tilde{h}_i\mathbf{\Delta }^{k_i+\kappa _i}}\mathbf{\Pi }^{l_i}
	\end{equation}
	is the channel matrix in the time domain, with $\mathbf{\Delta }$ being a constant $N\times N$ diagonal matrix defined by 
	\begin{equation} \label{eqn:def_Delta}
		\mathbf{\Delta }=\mathrm{diag}\left( \left[ e^{j\frac{2\pi}{N}\cdot 0},e^{j\frac{2\pi}{N}\cdot 1},\cdots ,e^{j\frac{2\pi}{N}\cdot \left( N-1 \right)} \right] \right) ,
	\end{equation}
	and $\mathbf{\Pi }$ being a $N\times N$ permutation matrix represented by
	\begin{equation}  \label{eqn:def_PI}
		\mathbf{\Pi }=\left[ \begin{matrix}
			0&		\cdots&		0&		1\\
			1&		\ddots&		0&		0\\
			\vdots&		\ddots&		\ddots&		\vdots\\
			0&		\cdots&		1&		0\\
		\end{matrix} \right]. 
	\end{equation}
	When a vector of length $N$ is multiplied by the permutation matrix $\mathbf{\Pi }$, its elements will be cyclic shifted forward by 1 element. For instance, for the signal vector $\boldsymbol{s}$,  we have $\mathbf{\Pi }\boldsymbol{s}=\left[ \left[ \boldsymbol{s} \right] _{N-1},\left[ \boldsymbol{s} \right] _0,\cdots ,\left[ \boldsymbol{s} \right] _{N-2} \right] ^T$. On the other hand, the effect of multiplying an $N$-length signal vector by matrix $\mathbf{\Delta }$ is equivalent to impose a Doppler shift on the signal vector with digital angular frequency $\frac{2\pi}{N}$. Therefore, as shown in (\ref{eqn:v_rcvsig}) and (\ref{eqn:v_chnnl}) , the received signal vector can be seen as the superposition of the cyclic-shifted, Doppler shifted, and attenuated versions of the transmit signal vector.  For the $i$-th path, the range of cyclic-shift, the digital angular frequency of Doppler-shift, and the equivalent path gain are $l_i$, $\frac{2\pi}{N}\left( k_i+\kappa _i \right)$, and $\tilde{h}_i$, respectively. 
	
	\section{Input-output Relation}
	In this section, given the channel coefficients of the physic paths, we first derive the general expression of the chirp-domain channel matrix, then propose a complexity-reduced method to compute the channel matrix.
	\subsection{General expression of the chirp-domain channel matrix}
	At the receiver, after obtaining the time domain signal vector $\boldsymbol{r}$, we use DFnT operation to convert it into a Fresnel domain vector as
	\begin{align}
		\boldsymbol{y}&=\mathbf{\Phi }\boldsymbol{r} \nonumber \\
		&=\mathbf{\Phi H\Phi }^{H}\boldsymbol{x}+\mathbf{\Phi }\boldsymbol{\omega } \nonumber \\
		&=\mathbf{H}_{\mathrm{eff}}\boldsymbol{x}+\tilde{\boldsymbol{\omega}},   \label{eqn:sysmodel-1}
	\end{align}
	where $\mathbf{H}_{\mathrm{eff}}=\mathbf{\Phi H\Phi }^H$ is the equivalent channel matrix in Fresnel domain, and $\tilde{\boldsymbol{\omega}}=\mathbf{\Phi }\boldsymbol{\omega }$ is the Fresnel domain complex AWGN vector.  Since DFnT is an orthogonal transform matrix, the noise vector $\tilde{\boldsymbol{\omega}}$ has zero mean and the same covariance matrix of $\boldsymbol{\omega }$, namely, $N_0\mathbf{I}_N$, which means that the elements of $\tilde{\boldsymbol{\omega}}$ are also IID random variables.
	
	To derive a simplified and intuitive expression of $\mathbf{H}_{\mathrm{eff}}$, we need to know the following fact that, the DFnT matrix $\mathbf{\Phi }$ (or the IDFnT matrix $\mathbf{\Phi }^H$) is a circulant matrix. Given any circulant matrix $
	\mathbf{A}$, it has the form
	\begin{equation}
		\mathbf{A}=\left[ \begin{matrix}
			a_0&		a_{N-1}&		\cdots&		a_1\\
			a_1&		a_0&		\cdots&		a_2\\
			\vdots&		\vdots&		\ddots&		\vdots\\
			a_{N-1}&		a_{N-2}&		\cdots&		a_0\\
		\end{matrix} \right], 
	\end{equation}
	where $\left[ \mathbf{A} \right] _{((m+1))_N,((n+1))_N}=\left[ \mathbf{A} \right] _{m,n}$, for $
	m,n=0,1,\cdots ,N-1$.
	From the definition of  $\mathbf{\Phi }$ and (\ref{eqn_discrete_chirp-2}), it is easy to verify that $
	\left[ \mathbf{\Phi } \right] _{((m+1))_N,((n+1))_N}=\left[ \mathbf{\Phi } \right] _{m,n}$. Thus, the matrix $\mathbf{\Phi }$ is circulant, and so does the matrix $\mathbf{\Phi }^H$.

	The following lemmas are also needed for the derivation of the channel matrix $\mathbf{H}_{\mathrm{eff}}$.
	\begin{lemma} \label{lemma1}
		For a given length-$N$ vector, the operation of Doppler shift after a cyclic shift can be realized by a cyclic shift after a Doppler shift. More specifically, the relation of the operation matrices can be expressed as
		\begin{equation} \label{eqn:exchange}
			\mathbf{\Delta \Pi }=e^{j\frac{2\pi}{N}}\mathbf{\Pi \Delta },
		\end{equation}
	\end{lemma}
	
	\emph{Proof}: See Appendix A.

	\begin{lemma} \label{lemma2}
		Given $N\times N$ matrices $\mathbf{\Phi}$, $\mathbf{\Delta }$, and $\mathbf{\Pi }$, defined in (\ref{eqn_SX}), (\ref{eqn:def_Delta}), and (\ref{eqn:def_PI}), respectively, we have
		\begin{equation}\label{eqn:lemma2}
			\mathbf{\Phi \Delta }^k\mathbf{\Phi }^{H}=e^{j\frac{\pi}{N}k^2}\mathbf{\Pi }^k\mathbf{\Delta }^k.
		\end{equation}
		Since, according to (\ref{eqn:lemma2}), for a length-$N$  chirp-domain signal vector $\boldsymbol{x}$, the vector $
		\mathbf{\Phi \Delta }^k\mathbf{\Phi }^{H}\boldsymbol{x}$ is equal to vector $e^{j\frac{\pi}{N}k^2}\mathbf{\Pi }^k\mathbf{\Delta }^k\boldsymbol{x}$, formula (\ref{eqn:lemma2}) can be explained as that, for a Fresnel domain vector, imposing a Doppler-shift of integer multiples of $2\pi/N$ on it in the discrete time domain, is equivalent to  imposing a cyclic-shift on it after a Doppler-shift in the Fresnel domain.
	\end{lemma}
	
	\emph{Proof}: See Appendix B.
	
	Substituting (\ref{eqn:v_chnnl}) into $\mathbf{H}_{\mathrm{eff}}=\mathbf{\Phi H\Phi }^H$, we have  
	\begin{align}
		\mathbf{H}_{\mathrm{eff}} &= \sum_{i=0}^{P-1}{\tilde{h}_i\mathbf{\Phi \Delta }^{\left( k_i+\kappa _i \right)}\mathbf{\Pi }^{l_i}\mathbf{\Phi }^{H}} \label{eqn:H_eff-01} \\
		&=\sum_{i=0}^{P-1}{\tilde{h}_i\mathbf{\Phi \Delta }^{\left( k_i+\kappa _i \right)}\mathbf{\Phi }^{H}\mathbf{\Pi }^{l_i}} \label{eqn:H_eff-02}
	\end{align}
	where the formula (\ref{eqn:H_eff-02}) is from (\ref{eqn:H_eff-01}) by the fact that both  
	$\mathbf{\Pi }^{l_i}$ and $\mathbf{\Phi }^H$ are circulant matrices, whose multiplying operation obeys the commutative law.
	
	With formula ({\ref{eqn:H_eff-02}}), we can directly calculate the chirp-domain channel matrix. However, there are a lot of matrix multiplication and matrix power operations, which will lead in high computational overhead.  On the other hand,  from this formula, we cannot directly see whether the matrix $\mathbf{H}_{\mathrm{eff}}$ is sparse. Hence, we need to continue the derivation of $\mathbf{H}_{\mathrm{eff}}$ based on it. In order to  make the result more intuitive, we first consider the case where the path Doppler shifts are integer multiples of $2\pi/N$, and then the case of fractional Doppler shifts.
	
	\emph{1) Case of integer Doppler factors}
	
	In this part, let us simplify the formula (\ref{eqn:H_eff-02}) under the assumption that all the path Doppler-shifts are integer multiples of $2\pi/N$ (digital angular frequency), namely, $\kappa_i=0$, for $i=0,1,\cdots,P-1$. Then, using  Lemma \ref{lemma2}, we can write formula (\ref{eqn:H_eff-02}) as
	\begin{align}
		\mathbf{H}_{\mathrm{eff}} &= \sum_{i=0}^{P-1}{\tilde{h}_i\mathbf{\Phi \Delta }^{k_i}\mathbf{\Phi }^H\mathbf{\Pi }^{l_i}} \nonumber \\
		&=\sum_{i=0}^{P-1}{\tilde{h}_ie^{j\frac{\pi}{N}{k_i}^2}\mathbf{\Pi }^{k_i}\mathbf{\Delta }^{k_i}\mathbf{\Pi }^{l_i}} \label{eqn:H_eff-03}\\
		&=\sum_{i=0}^{P-1}{\tilde{h}_ie^{-j\frac{\pi}{N}{k_i}^2}\mathbf{\Delta }^{k_i}\mathbf{\Pi }^{l_i+k_i}}, \label{eqn:H_eff-04}
	\end{align}
	where the formula (\ref{eqn:H_eff-04}) is derived from (\ref{eqn:H_eff-03}) by using equation $
	\mathbf{\Pi }^{k_i}\mathbf{\Delta }^{k_i}=e^{-j\frac{2\pi}{N}k_{i}^{2}}\mathbf{\Delta }^{k_i}\mathbf{\Pi }^{k_i}
	$, obtained from Lemma \ref{lemma1}. 
	
	Now, let us omit the channel noise to intuitively explain the expression of the received Fresnel domain signal in OCDM  over MLMD channels. As shown in (\ref{eqn:H_eff-04}), given a Fresnel domain signal vector $\boldsymbol{x}$, when the path Doppler-shifts factors are integers, we can regard the received signal vector as the superposition of copies of the transmit vector $\boldsymbol{x}$ after cyclic shift and Doppler shift in the Fresnel domain. Specifically, for the $i$-th physic path, with path gain $h_i$, time-domain delay $l_i$ and Doppler shift with frequency $2\pi k_i/N$, its effect on the transmit vector $\boldsymbol{x}$ can be seen as a  chirp-domain Doppler shift with frequency $2\pi k_i/N$ after a cyclic shift by $l_i+k_i$ elements, with attenuation coefficient $\tilde{h}_ie^{-j\frac{\pi}{N}{k_i}^2}$.
	
	It can be inferred from (\ref{eqn:H_eff-04}) that $\mathbf{H}_{\mathrm{eff}}$ is a sparse matrix, because the permutation matrix $\mathbf{\Pi}^{l_i+k_i}$ is sparse, multiplying it by a diagonal matrix $\mathbf{\Delta }^{k_i}$ keeps the sparsity, and the path number   $P$ is generally far less than the chirp number $N$. The sparsity of $\mathbf{H}_{\mathrm{eff}}$ can be exploited to design complexity reduced  detector of transmit symbols, which is presented in section \ref{sec:Detector} of this paper.

	\emph{2) Case of fractional Doppler factors}
	
	In practical applications, the Doppler shifts of channel paths may be fractional multiples of $2\pi/N$. Namely, the value of  ${\kappa _i}$ in (\ref{eqn:H_eff-02}) may not be zero but a fractional number in $\left(-0.5, 0.5 \right]$. For the $i$th path, let us define a vector 
	\begin{equation} \label{eqn:fracDoppler-001}
		\boldsymbol{v}_i=\left[ e^{j\frac{2\pi}{N}\kappa _i\cdot 0},e^{j\frac{2\pi}{N}\kappa _i\cdot 1},\cdots ,e^{j\frac{2\pi}{N}\kappa _i\cdot \left( N-1 \right)} \right] ^T,
	\end{equation}
	with which $\mathbf{\Delta }^{\kappa _i}$ can be expressed as
	\begin{equation} \label{eqn:fracDoppler-002}
		\mathbf{\Delta }^{\kappa _i} = \mathrm{diag}\left( \boldsymbol{v}_i \right). 
	\end{equation}
	On the other hand, in the dimension-$N$ linear complex vector space, $\mathbb{C} ^N$, we define a set of complex vectors $\mathcal{V} =\left\{ \boldsymbol{\vartheta }_m| m=-\frac{N}{2},-\frac{N}{2}+1,\cdots ,0,\cdots,\frac{N}{2}-1 \right\} $, where
	\begin{equation} 
		\boldsymbol{\vartheta }_m\,\,=\,\,\left[ e^{j\frac{2\pi}{N}m\cdot 0},e^{j\frac{2\pi}{N}m\cdot 1},\cdots ,e^{j\frac{2\pi}{N}m\cdot \left( N-1 \right)} \right] ^T.
	\end{equation}
	It is easy to verify that the vector space  $\mathbb{C} ^N$ can be spanned by the vectors of $\mathcal{V}$. Namely, $\mathcal{V}$ is a basis of  $\mathbb{C} ^N$. Then, the vector $\boldsymbol{v}_i$ can be expressed by a linear combination of those basis vectors as
	\begin{equation} 
		\boldsymbol{v}_i=\sum_{m=-N/2}^{N/2-1}{\lambda _{i,m}\boldsymbol{\vartheta }_m} \label{eqn:fracDoppler-003}
	\end{equation}
	with
	\begin{align}
		\lambda _{i,m}&=\frac{\boldsymbol{\vartheta }_{m}^{H}\boldsymbol{v}_i}{\boldsymbol{\vartheta }_{m}^{H}\boldsymbol{\vartheta }_m} \nonumber \\
		&=\frac{1}{N}\sum_{n=0}^{N-1}{e^{j\frac{2\pi}{N}\kappa _in}}e^{-j\frac{2\pi}{N}mn} \label{eqn:fracDoppler-004} \\
		&=\frac{1}{N}\frac{e^{j2\pi \kappa _i}-1}{e^{j\frac{2\pi}{N}\left( \kappa _i-m \right)}-1}, \label{eqn:fracDoppler-005} 
	\end{align}
	where the formula (\ref{eqn:fracDoppler-005}) is obtained by summing the geometric sequence in (\ref{eqn:fracDoppler-004}). It should be noted that, for the $i$-th path, the value of $\lambda_{i,m}$ decreases rapidly as the absolute value of $m$ increases, because the modulus of the denominator, $e^{j\frac{2\pi}{N}\left( \kappa _i-m \right)}-1$, is close to zero  when $\left| m \right|$ is a small integer, and it increases with $\left| m \right|$.  Therefore, according to (\ref{eqn:fracDoppler-003}),  we can approximate $\boldsymbol{v}_i$ as 
	\begin{equation} \label{eqn:fracDoppler-006}
		\boldsymbol{v}_i\approx \sum_{m=-M_i}^{M_i}{\lambda _{i,m}\boldsymbol{\vartheta }_m},
	\end{equation}
	where the value of $M_i$, is usually far less than $N/2$, and depends on the desired accuracy of the approximation and the value of $\kappa _i$. As a special case, when $\kappa _i = 0$, $M_i$ should be set to 0, because we can directly write formula (\ref{eqn:fracDoppler-006}) as $\boldsymbol{v}_i=\boldsymbol{\vartheta }_0$ in this case.
	
	From (\ref{eqn:H_eff-02}), (\ref{eqn:fracDoppler-002}), (\ref{eqn:fracDoppler-003}), and (\ref{eqn:fracDoppler-006}), the Fresnel domain channel matrix can be approximated as  $
	\tilde{\mathbf{H}}_{\mathrm{eff}}\approx \mathbf{H}_{\mathrm{eff}}$, with
	
	\begin{align}
		\tilde{\mathbf{H}}_{\mathrm{eff}}& = \sum_{i=0}^{P-1}{\sum_{m=-M_i}^{M_i}{\tilde{h}_i\lambda _{i,m}e^{-j\frac{\pi}{N}{k_i}^2}\mathbf{\Phi \Delta }^m\mathbf{\Phi }^H\mathbf{\Delta }^{k_i}\mathbf{\Pi }^{l_i+k_i}}}\nonumber \\
		&= \sum_{i=0}^{P-1}{\sum_{m=-M_i}^{M_i}{\tilde{h}_i\lambda _{i,m}e^{-j\frac{\pi}{N}{k_i}^2}e^{j\frac{\pi}{N}m^2}\mathbf{\Pi }^m\mathbf{\Delta }^m\mathbf{\Delta }^{k_i}\mathbf{\Pi }^{l_i+k_i}}}    \label{eqn:fracDoppler-007} \\
		&= \sum_{i=0}^{P-1}{\sum_{m=-M_i}^{M_i}{\tilde{h}_i\lambda _{i,m}e^{-j\frac{\pi}{N}\left( k_i+m \right) ^2}\mathbf{\Delta }^{k_i+m}\mathbf{\Pi }^{k_i+m+l_i}}}, \label{eqn:fracDoppler-008} 
	\end{align}
	where the formulas  (\ref{eqn:fracDoppler-007}) and (\ref{eqn:fracDoppler-008}) are derived from their previous ones by using Lemma \ref{lemma2} and Lemma \ref{lemma1} respectively
	
	So far, as shown in  (\ref{eqn:fracDoppler-008}), we complete the derivation of the general channel matrix. Based on the explanation of (\ref{eqn:H_eff-04}), the resulted formula can be explained as follows by a intuitive way. The effect of path $i$ with fractional Doppler factor $k_i+\kappa _i$ can be approximately realized by $2M_i+1$ virtual paths with integer Doppler factors. And the channel matrix $\tilde{\mathbf{H}}_{\mathrm{eff}}$ can be seen as the superposition of a group of permutation matrices with non-zero elements being weighted, while each additive term corresponds to a virtual path.

	\subsection{Complexity-reduced computation of the chirp-domain channel matrix}
	Although the form of formula (\ref{eqn:fracDoppler-008}) is very simple and intuitive, we do not directly use it to calculate the channel matrix $\tilde{\mathbf{H}}_{\mathrm{eff}}$ at the receiver. Because there are still some multiplication and exponentiation on $N\times N$ matrices, which will result in high computational overhead.  In this subsection, based on it, we propose a complexity-reduced computation of the channel matrix.

	It can be seen from (\ref{eqn:fracDoppler-008}) that, the $m$-th virtual path of physic path $i$  has the following parameters, attenuation coefficient $\tilde{h}_i\lambda _{i,m}e^{-j\frac{\pi}{N}\left( k_i+m \right) ^2}$, Doppler shift $k_i+m$, and time delay $l_i$. 
	It should be noted that, the exponent of the cyclic-shift matrix corresponding to the virtual path is $k_i+m+l_i$, which means that the symbol shift in Fresnel domain are resulted by the time-domain delay and the Doppler shift together. Here, we define a new parameter, called \emph{chirp-shift}, to describe the cyclic-shifts of the chirp symbols resulted by a virtual path. Apparently, different virtual paths may have the same chirp-shifts. In (\ref{eqn:fracDoppler-008}), all the virtual paths of the same chirp-shift can be seen as being merged into an equivalent path expressed by a unique cyclic-shift matrix. Let us call the equivalent path as \emph{logical path} in the rest of this paper, and rewrite the channel matrix of (\ref{eqn:fracDoppler-008}) by logical paths as 
	\begin{align}
		\tilde{\mathbf{H}}_{\mathrm{eff}}&= \sum_{\ell =0}^{L-1}{\left( \sum_{\left( i,m \right) \in \mathcal{A} _{\ell}}{\tilde{h}_i\lambda _{i,m}e^{-j\frac{\pi}{N}\left( k_i+m \right) ^2}\mathbf{\Delta }^{k_i+m}} \right) \mathbf{\Pi }^{d_{\ell}}} \nonumber \\
		&= \sum_{\ell =0}^{L-1}{\mathrm{diag}\left( \breve{\boldsymbol{h}}_\ell \right) \mathbf{\Pi }^{d_{\ell}}}, \label{eqn:MP-Heff}
	\end{align}
	with 
	\begin{equation}
		\breve{\boldsymbol{h}}_\ell \triangleq \sum_{\left( i,m \right) \in \mathcal{A} _{\ell}}{\tilde{h}_i\lambda _{i,m}e^{-j\frac{\pi}{N}\left( k_i+m \right) ^2}\breve{\boldsymbol{\vartheta}}_{k_i+m}}, \label{eqn:h_l_1}
	\end{equation}
	\begin{equation}
		\breve{\boldsymbol{\vartheta}}_n\,\,=\,\,\left[ e^{j\frac{2\pi}{N}n\cdot 0},e^{j\frac{2\pi}{N}n\cdot 1},\cdots ,e^{j\frac{2\pi}{N}n\cdot \left( N-1 \right)} \right] ^T,n\in z,
	\end{equation}
	where $d_\ell$ is the chirp-shift of the $\ell$-th logical path, and $d_{\ell_1} \ne d_{\ell_2}$, for any given $\ell_1$, $\ell_2$ with $\ell_1 \ne \ell_2$, $\mathcal{A} _l$ is the set of indices of the virtual paths contributed to the $l$-th logical path, defined as
	\begin{align}
		\mathcal{A} _{\ell}= & \left\{ \left( i,m \right) |l_i+k_i+m=d_{\ell}, \right. \nonumber \\
		& \left. i=0,\cdots ,P-1,m=-M_i,\cdots ,M_i \right\}.
	\end{align}
	It can be observed from ({\ref{eqn:MP-Heff}}) that, for a given transmit vector $\boldsymbol{x}$ , via the $\ell$-th logical path , the transmit symbols are first cyclic shifted by $d_\ell$ elements, then weighted element-by-element by weights presented in vector $\breve{\boldsymbol{h}}_\ell$, 
	at last superimposed onto the received signal vector. Based on this, we can infer from  ({\ref{eqn:MP-Heff}}) that the nonzero entries of $\tilde{\mathbf{H}}_{\mathrm{eff}}$ can be directly expressed by
	\begin{align}
		&\left[ \left[ \tilde{\mathbf{H}}_{\mathrm{eff}} \right] _{\left[ \boldsymbol{g}_\ell \right] _0,0}, \left[ \tilde{\mathbf{H}}_{\mathrm{eff}} \right] _{\left[ \boldsymbol{g}_\ell \right] _1,1}, \cdots ,\left[ \tilde{\mathbf{H}}_{\mathrm{eff}} \right] _{\left[ \boldsymbol{g}_\ell \right] _{N-1},N-1} \right] ^T = \breve{\boldsymbol{h}}_\ell,\nonumber \\
		& \quad l = 0,1,\cdots,L-1,    \label{eqn:computeH}
	\end{align}
	where 
	\begin{equation}
		\boldsymbol{g}_\ell=\left( \left( \left[ 0,1,\cdots ,N-1 \right] ^T+d_\ell \right) \right) _N \label{eqn:h_l_2}
	\end{equation}
	is the indices vector whose $n$-th element denotes the index of the received symbol superimposed by the $n$-th transmit symbol via the $\ell$-th logical path. \par
	Therefore, in practical applications, with (\ref{eqn:computeH}), we can achieve low-complexity computation of  $\tilde{\mathbf{H}}_{\mathrm{eff}}$ by directly calculating its nonzero elements, without matrix multiplication or matrix exponentiation.

	\section{Message Passing Based Detector} \label{sec:Detector}
	In this section, under the assumption that the path coefficients of the physic paths are perfectly estimated by the receiver, we propose an iterative detector using the message passing algorithm, which can exploit the sparsity of the  chirp-domain channel matrix $\tilde{\mathbf{H}}_{\mathrm{eff}}$. 
	
	Considering the influence of the approximation of channel matrix $\mathbf{H}_{\mathrm{eff}}$, based on (\ref{eqn:sysmodel-1}), we rewrite the chirp-domain system model as 
	\begin{equation}
		\boldsymbol{y}=\tilde{\mathbf{H}}_{\mathrm{eff}}\boldsymbol{x}+\breve{\boldsymbol{\omega}}
	\end{equation}
	where $\breve{\boldsymbol{w}}$ is the additive noise vector, composed by two parts, the chirp-domain channel AWGN $
	\tilde{\boldsymbol{\omega}}$ (see in (\ref{eqn:sysmodel-1})) and the noise vector caused by channel approximation. For the sake of simplicity, we assume that the elements of the second part of noise are also independent and identically-distributed AWGN. 
	
\begin{figure*}[!ht]
		\includegraphics[width=18cm]{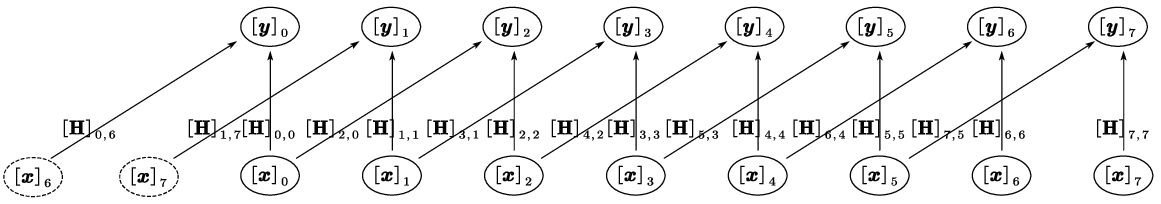}
		\caption{\small The chirp-domain input-output relation of an OCDM system of 8 orthogonal chirps, while the channel has 2 logical paths, with chirp-shifts 0 and 2 respectively.}
		\label{diagram_channelExample}
	\end{figure*}

	It can be inferred from (\ref{eqn:MP-Heff}) that each row of channel matrix $\tilde{\mathbf{H}}_{\mathrm{eff}}$ has only $L$ non-zero elements, and so does each column. Namely, for an arbitrary transmit symbol, it will be delivered to $L$ elements of the received vector. On the other hand, in addition to the noise, each received symbol is superimposed by $L$ transmit symbols, which interfere with each other. For example, in Fig. (\ref{diagram_channelExample}), we illustrate the relation of the input and output symbols of an OCDM system with 8 orthogonal chirps, while the channel has 2 logical paths.
	
	Let us denote the indexes of the transmit symbols contributing to $\left[ \boldsymbol{y} \right] _n$ and the indexes of the received symbols contributed by  $\left[ \boldsymbol{x} \right] _n$ as vectors $
	\boldsymbol{b}_n$ and $\boldsymbol{q}_n$, respectively. According to (\ref{eqn:MP-Heff}), we have 
	
	\begin{equation}
		\boldsymbol{b}_n=\left( \left( n-\left[ d_0,d_1,\cdots ,d_{L-1} \right] \right) \right) _N, 
	\end{equation}
	and
	\begin{equation}
		\boldsymbol{q}_n=\left( \left( n+\left[ d_0,d_1,\cdots ,d_{L-1} \right] \right) \right) _N, 
	\end{equation}
	with $\left[ \boldsymbol{b}_n \right] _{\ell}$ and $\left[ \boldsymbol{q}_n \right] _{\ell}$ both being related to the $\ell$-th logical path.
	
	The $n$-th element of the received vector $\boldsymbol{y}$, $\left[ \boldsymbol{y} \right] _n$, can be seen as the observation of random variable  $\left[ \boldsymbol{x} \right] _{\left[ \boldsymbol{b}_n \right] _{\ell}}$, $\ell = 0,1,\cdots,L-1$, with relation expressed by
	\begin{align}
		\left[ \boldsymbol{y} \right] _n &=
		{\left[ \tilde{\mathbf{H}}_{\mathrm{eff}} \right] _{n,}}_{\left[ \boldsymbol{b}_n \right] _{\ell}}\cdot \left[ \boldsymbol{x} \right] _{\left[ \boldsymbol{b}_n \right] _{\ell}} \nonumber \\
		&+\mathop {\underbrace{\sum_{\begin{array}{c}
						i=0\\
						i\ne \ell \\
				\end{array}}^{L-1}{{\left[ \tilde{\mathbf{H}}_{\mathrm{eff}} \right] _{n,}}_{\left[ \boldsymbol{b}_n \right] _i}\cdot \left[ \boldsymbol{x} \right] _{\left[ \boldsymbol{b}_n \right] _i}+\left[ \breve{\boldsymbol{\omega}} \right] _n}}} \limits_{\left[ \mathbf{W} \right] _{n,\left[ \boldsymbol{b}_n \right] _{\ell}}}, \label{ISI_model}
	\end{align}
	where $\mathbf{W}$ is a sparse matrix, with $\left[ \mathbf{W} \right] _{n,\left[ \boldsymbol{b}_n \right] _{\ell}}$ denoting the interference and noise  superimposed on transmit symbol $\left[ \boldsymbol{x} \right] _{\left[ \boldsymbol{b}_n \right] _{\ell}}$,  contained by $\left[ \boldsymbol{y} \right] _n$. For the sake of simplicity, we assume that all the elements in  $\boldsymbol{x}$ are independent of each other. And we also assume that the random variable $\left[ \mathbf{W} \right] _{n,\left[ \boldsymbol{b}_n \right] _{\ell}}$ follows a Gaussian distribution. According to (\ref{ISI_model}), we know the following two facts.
	
	\emph{1)} Provided the probability distribution of $\boldsymbol{x}$, we can calculate the expectation and variance of $ \left[ \mathbf{W} \right] _{n,\left[ \boldsymbol{b}_n \right]_{\ell}}$, with $n=0, \cdots, N-1$, $\ell=0, 1, \cdots, L-1$. 
	
	\emph{2)} Provided the expectation and variance of $ \left[ \mathbf{W} \right] _{n,\left[ \boldsymbol{b}_n \right]_{\ell}}$, with $n=0, \cdots, N-1$, $\ell=0, 1, \cdots, L-1$,  we can calculate the posterior probability of transmit symbol $\left[ \boldsymbol{x} \right] _{\left[ \boldsymbol{b}_n \right] _{\ell}}$ based on observation $\boldsymbol{y} $.
	
	\begin{figure}[!htbp]
		\centering
		\includegraphics[width=8cm]{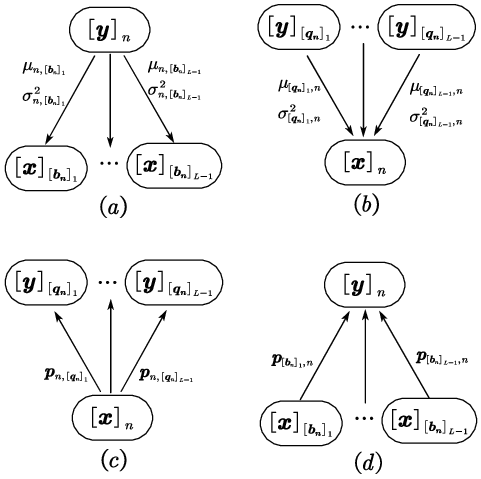}
		\caption{\small Messages in factor graph: (a) massages sent from an observation node to $L$ variable nodes, (b) massages got by an variable node from $L$ observation nods, (c) messages sent from an variable node to $L$ observation nodes, (d) messages got by an observation node from $L$ variable nodes.}
		\label{fig:factorGraph}
	\end{figure}
	
	The basic idea of the message passing detector is to iteratively calculate the posterior probability of the transmit symbols in $\boldsymbol{x}$, based on the above two facts\cite{OTFS-HongYi}                                                                                                                                                                                                                                                                                                                                                                                                                                                                                                                                                                                                                                                                                                                                                                                                                                                                                                                                                                                                                                                                                                                                                                                                                                                                                                                                                                                                                                              . In Fig. \ref{fig:factorGraph}, we illustrate the sparsely connected factor graph with $N$ observation nodes (elements of $\boldsymbol{y}$)  and $N$ variable nodes (elements of $\boldsymbol{x}$). As shown in subfigure (a), the messages sent from the observation nodes $\left[ \boldsymbol{y} \right] _n$ to $\left[ \boldsymbol{x} \right] _{\left[ \boldsymbol{b}_{\boldsymbol{n}} \right] _{\ell}}$ are the expectation and variance of $\left[ \mathbf{W} \right] _{n,\left[ \boldsymbol{b}_n \right] _{\ell}}$, defined in (\ref{ISI_model}). As shown in subfigure (c), the message sent from variable node $\left[ \boldsymbol{x} \right] _n$ to $\left[ \boldsymbol{y} \right] _{\left[ \boldsymbol{p}_{\boldsymbol{n}} \right] _{\ell}}$ is the probability mass function (pmf) $
	\boldsymbol{p}_{n,\left[ \boldsymbol{q}_n \right] _{\ell}}=\left\{ p_{n,\left[ \boldsymbol{q}_n \right] _{\ell}}\left( \alpha _m \right) |\alpha _m\in \mathcal{X} \right\} $, with
	$ p_{n,\left[ \boldsymbol{q}_n \right] _{\ell}}\left( \alpha _m \right)$ being the posterior probability of the event $\left[ \boldsymbol{x} \right] _n=\alpha _m$ calculated based on the values of the observation nodes and the messages that have been obtained.  Sub-figures (b) and (d) show that, either each variable node or each observation node receives $L$ messages from $L$ counterparts related by the $L$ logical paths.

	\begin{algorithm}[!t]
		\caption{MP Algorithm for OCDM Symbol Detection}
		\label{alg1}
		\begin{algorithmic}[1]
			\REQUIRE 
			$\boldsymbol{y}$ (received chirp-domain vector), \\
			$\tilde{\mathbf{H}}_{\mathrm{eff}}$ (chirp-domain channel matrix),\\
			$\boldsymbol{b}_n$ and $\boldsymbol{q}_n$, $n=0,1,\cdots,N-1$ (indices vectors), \\
			$\sigma _{o}^{2}$ (variance of elements of IID AWGN $\breve{\boldsymbol{\omega}}$). \\
			$I_{\max}$ (the allowed number of iterations).
			
			\ENSURE  
			$\hat{\boldsymbol{x}}$ (estimation of the transmit symbols) \\
			
			\STATE \textbf{Initiation:} $p_{n,\left[ \boldsymbol{q}_n \right] _{\ell}}\left( \alpha _m \right) \leftarrow 1/M$, with $n=0,\cdots,N-1$, \\
			$\ell=0, \cdots, L-1$, and $m=0,\cdots,M-1$. \\ 
			$i \leftarrow 1$ (iteration count), \\
			$\eta \leftarrow 0$ (convergence indicator), \\
			$\eta _{\max} \leftarrow \eta$ (maximum $\eta$ in previous iterations).\\
			
			\WHILE{$\left( i\leqslant I_{\max} \right) \mathrm{and} \left( \eta <1 \right)$}
			\STATE Compute messages $\mu _{n,\left[ \boldsymbol{b}_n \right] _{\ell}}$ and $\sigma _{n,\left[ \boldsymbol{b}_n \right] _{\ell}}^{2}$ by (\ref{eqn:mean_awgn}) and (\ref{eqn:var_awgn}), respectively, for  $n=0,\cdots,N-1$,
			$\ell=0, \cdots, L-1$,
			\STATE Update messages $p_{n,\left[ \boldsymbol{q}_n \right] _{\ell}}\left( \alpha _m \right) \leftarrow \varDelta \tilde{p}_{n,\left[ \boldsymbol{q}_n \right] _{\ell}}\left( \alpha _m \right)$ \\
			$+\left( 1-\varDelta \right) p_{n,\left[ \boldsymbol{q}_n \right] _{\ell}}\left( \alpha _m \right)$, with $
			\tilde{p}_{n,\left[ \boldsymbol{q}_n \right] _{\ell}}\left( \alpha _m \right)$ calculated by (\ref{eqn:tilt_p}), (\ref{eqn:breve_p}), and (\ref{eqn:awgn_pdf}), for  $n=0,\cdots,N-1$,
			$\ell=0, \cdots, L-1$, $m=0,\cdots,M-1$.
			
			\STATE  Compute convergence indicator $\eta$ by formulas (\ref{eqn:convergence_factor}), (\ref{eqn:combinedProb_xn}), and (\ref{eqn:raw_combinedProb_xn}).
			\IF{$\eta >\eta _{\max}$}
			\STATE $\eta _{\max}\gets \eta$ 
			\STATE Update $\hat{\boldsymbol{x}}$,  the estimation of $\boldsymbol{x}$,  by (\ref{eqn:x_estimation}).
			\ELSIF{$\eta <\eta _{\max}-\epsilon$} 
			\STATE \textbf{break}
			\ENDIF
			\STATE $i\leftarrow i+1$
			\ENDWHILE
			\RETURN $\hat{\boldsymbol{x}}$
			
		\end{algorithmic}
	\end{algorithm}
	
	The MP detecting algorithm is presented in \textbf{Algorithm} \ref{alg1}. In the algorithm,  the message sent by the variable nodes, denoted by $p_{n,\left[ \boldsymbol{q}_n \right] _{\ell}}\left( \alpha _m \right) $, are initiated with $1/M$, for  $n=0,\cdots,N-1$, $\ell=0, \cdots, L-1$, and $m=0,\cdots,M-1$. During the iteration, the messages sent from the observation nodes are calculated by 
	
	\begin{equation}
		\mu _{_{n,\left[ \boldsymbol{b}_n \right] _{\ell}}}=\sum_{\begin{array}{c}
				i=0\\
				i\ne \ell\\
		\end{array}}^{L-1}{\sum_{m=0}^{M-1}{p_{\left[ \boldsymbol{b}_n \right] _i,n}\left( \alpha _m \right) {\left[ \tilde{\mathbf{H}}_{\mathrm{eff}} \right] _{n,}}_{\left[ \boldsymbol{b}_n \right] _i}\alpha _m}}, \label{eqn:mean_awgn}
	\end{equation}
	and
	\begin{align}
		\sigma _{n,\left[ \boldsymbol{b}_n \right] _{\ell}}^{2}=&\sum_{\begin{array}{c}
				i=0\\
				i\ne \ell\\
		\end{array}}^{L-1}{\left( \sum_{m=0}^{M-1}{p_{\left[ \boldsymbol{b}_n \right] _i,n}\left( \alpha _m \right) \left| {\left[ \tilde{\mathbf{H}}_{\mathrm{eff}} \right] _{n,}}_{\left[ \boldsymbol{b}_n \right] _i} \right|^2\left| \alpha _m \right|^2}
			\right. }  \nonumber\\
		& \left. -\left| \sum_{m=0}^{M-1}{p_{\left[ \boldsymbol{b}_n \right] _i,n}\left( \alpha _m \right) \alpha _m{\left[ \tilde{\mathbf{H}}_{\mathrm{eff}} \right] _{n,}}_{\left[ \boldsymbol{b}_n \right] _i}} \right|^2\right)+\sigma_{o,}^{2}, \label{eqn:var_awgn}
	\end{align}
	for $n=0,\cdots,N-1$, $\ell=0, \cdots, L-1$. Based on the latest messages obtained from the observation nodes, the variable nodes update their messages by replacing  $p_{n,\left[ \boldsymbol{q}_n \right] _{\ell}}\left( \alpha _m \right)$ with $\varDelta \tilde{p}_{n,\left[ \boldsymbol{q}_n \right] _{\ell}}\left( \alpha _m \right)
	+\left( 1-\varDelta \right) p_{n,\left[ \boldsymbol{q}_n \right] _{\ell}}\left( \alpha _m \right)$, where $\varDelta$ is the \emph{damping factor} , a positive constant less than 1, used to control the convergence speed, and the term $\tilde{p}_{n,\left[ \boldsymbol{q}_n \right] _{\ell}}\left( \alpha _m \right)$ is calculated by
	\begin{equation}
		\tilde{p}_{n,\left[ \boldsymbol{q}_n \right] _{\ell}}\left( \alpha _m \right) =\frac{\breve{p}_{n,\left[ \boldsymbol{q}_n \right] _{\ell}}\left( \alpha _m \right)}{\sum_{k=0}^{M-1}{\breve{p}_{n,\left[ \boldsymbol{q}_n \right] _{\ell}}\left( \alpha _k \right)}}, \label{eqn:tilt_p}
	\end{equation}
	where
	\begin{equation}
		\breve{p}_{n,\left[ \boldsymbol{q}_n \right] _{\ell}}\left( \alpha _m \right) =\prod_{i=0,i\ne \ell}^{L-1}{\frac{p\left( \left[ \boldsymbol{y} \right] _{_{\left[ \boldsymbol{q}_n \right] _i}}|\left[ \boldsymbol{x} \right] _n=\alpha _m \right)}{\sum_{k=0}^{M-1}{p\left( \left[ \boldsymbol{y} \right] _{_{\left[ \boldsymbol{q}_n \right] _i}}|\left[ \boldsymbol{x} \right] _n=\alpha _k \right)}}}, \label{eqn:breve_p}
	\end{equation}
	with
	\begin{align}
		&p\left( \left[ \boldsymbol{y} \right] _{_{\left[ \boldsymbol{q}_n \right] _i}}|\left[ \boldsymbol{x} \right] _n=\alpha _m \right)= \nonumber\\
		&\exp \left( \frac{-\left| \left[ \boldsymbol{y} \right] _{\left[ \boldsymbol{q}_n \right] _i}-\left[ \tilde{\mathbf{H}}_{\mathrm{eff}} \right] _{\left[ \boldsymbol{q}_n \right] _i,n}\left[ \boldsymbol{x} \right] _n-\mu _{\left[ \boldsymbol{q}_n \right] _i,n} \right|^2}{\sigma _{\left[ \boldsymbol{q}_n \right] _i,n}^{2}} \right). \label{eqn:awgn_pdf}
	\end{align}
	At each iteration, after the messages are updated, the convergence factor $\eta$ is computed by
	\begin{equation}
		\eta =\frac{1}{N}\sum_{n=0}^{N-1}{\mathbb{I} \left( \mathop {\max } \limits_{\alpha _m\in \mathcal{X}}p_n\left( \alpha _m \right) \geqslant \gamma \right)}, \label{eqn:convergence_factor}
	\end{equation}
	where $\gamma$ is a constant less than 1 and close to 1; the value of $\mathbb{I} \left( \cdot \right)$ is $1$ when the expression in the brackets is true, and 0 otherwise; and $p_n\left( \alpha _m \right)$ is the posterior probability of $\left[ \boldsymbol{x} \right] _n$ given all the $L$ corresponding observation nodes, calculated by 
	\begin{equation}
		p_n\left( \alpha _m \right) =\frac{\hat{p}_n\left( \alpha _m \right)}{\sum_{k=0}^{M-1}{\hat{p}_n\left( \alpha _k \right)}}, \label{eqn:combinedProb_xn}
	\end{equation}
	with
	\begin{equation}
		\hat{p}_n\left( \alpha _m \right) =\prod_{i=0}^{L-1}{\frac{p\left( \left[ \boldsymbol{y} \right] _{_{\left[ \boldsymbol{q}_n \right] _i}}|\left[ \boldsymbol{x} \right] _n=\alpha _m \right)}{\sum_{k=0}^{M-1}{p\left( \left[ \boldsymbol{y} \right] _{_{\left[ \boldsymbol{q}_n \right] _i}}|\left[ \boldsymbol{x} \right] _n=\alpha _k \right)}}} \label{eqn:raw_combinedProb_xn}.
	\end{equation}
	Note that the terms in the right hand of formula ($\ref{eqn:raw_combinedProb_xn}$) can be calculated by (\ref{eqn:awgn_pdf}).  After the convergence factor is updated, it will be compared with its historical maximum value $\eta _{\max}$.
	If the current convergence factor $\eta$ is greater than $\eta _{\max}$, $\eta _{\max}$ is firstly updated with  $\eta$. Then the estimation of the transmit symbol vector $\hat{\boldsymbol{x}}$ is updated with
	\begin{equation}
		\left[ \hat{\boldsymbol{x}} \right] _n=\mathop {\mathrm{arg}\max} \limits_{\alpha _m\in \mathcal{X}}\,\,p_n\left( \alpha _m \right). \label{eqn:x_estimation}
	\end{equation}
	It should be noted that, the iteration of the algorithm will be terminated when any of the following three conditions is met.
	\begin{itemize}
		\item The number of iterations reaches the maximum allowed value $I_{\max}$.
		\item The convergence factor $\eta$ reaches 1.
		\item The convergence factor $\eta$ regresses obviously with $\eta <\eta _{\max}-\epsilon$, where $\epsilon$ is a small positive number defining the regression tolerance of the convergence factor. 
	\end{itemize}
	
	\section{Simulation results and discussion}
	
	In this section, we use Monte Carlo simulation to verify the derived chirp-domain channel matrix and evaluate the proposed MP detector. Under the assumptions of perfect and imperfect channel state information (CSI) at the receiver, the bit error rate (BER) performance of OCDM and OFDM are tested under LTV channels with multi-lags and multi-Dopplers. Without loss of generality, the influence of CP has been taken into account in $E_b/N_0$ calculation in all the simulations. 
	
	At first, we consider a terrestrial radio OCDM system for uncoded transmissions. The carrier frequency, the signal bandwidth,  the number of chirps, the symbol constellation for data bits mapping, the duration of symbol with CP, the guard interval and the UE speed are listed  in Table \ref{table-simParas}. Extended Vehicular A (EVA) model \cite{EVA-Model} is adopted for setting the delay-power profile of the simulation channel, which is shown in Table \ref{table-radio-DelayPower}.  Corresponding to the UE speed and the carrier frequency, the maximum path Doppler shift $v_{\max}$ is 2315 Hz, whose normalized value is 0.0386 with the normalizing frequency interval $\Delta f$. The Doppler-shift of the  $i$-th path is $v_i = v_{\max} \cos{\theta}$, where $\theta$ is a random number uniformly distributed in $\left[ -\pi /2, \pi /2 \right] $. Therefore, the integer parts of all the path Doppler-shifts are zeros, and their fractional parts are far less than 0.5. In this case, Equation  (\ref{eqn:fracDoppler-006}) can obtain a good approximation even if    $M_i$ is set to a small number. In the simulation, we set $M_i=5$ for all the paths. The damping factor $\Delta$ and the allowed number of iterations of the MP detector are set to be 0.6 and 20 respectively. 
	
\begin{figure}[!t]
		\centering
		\includegraphics[width=8.8cm]{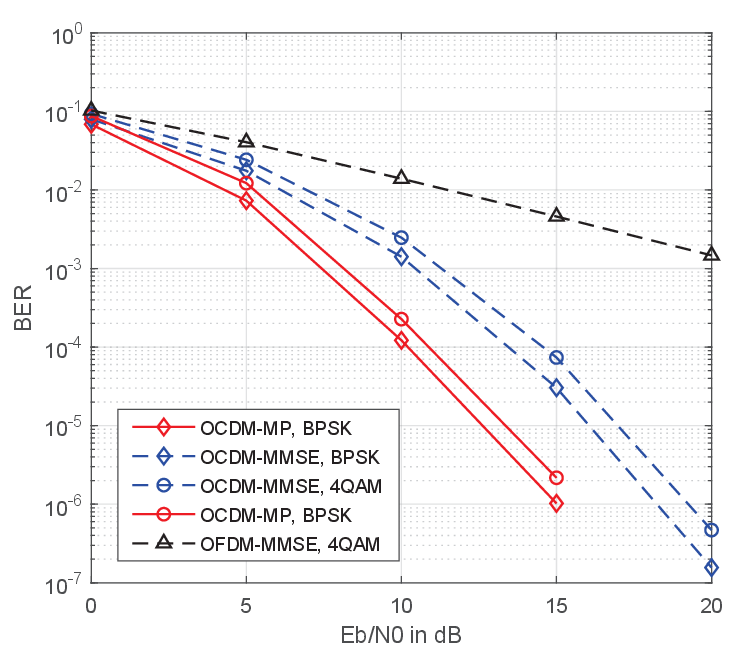}
		\caption{BER performance comparison between OCDM-MP, OCDM-MMSE, and OFDM-MMSE under EVA channel model with UE speed 500 kmph.}\label{fig:radio_BER}	
	\end{figure}

	Fig. \ref{fig:radio_BER} depicts the comparison of BER performance between  OCDM with MP receiver (OCDM-MP), OCDM with MMSE receiver (OCDM-MMSE), and OFDM with MMSE receiver (OFDM-MMSE), under EVA channel model with UE speed being 500 kilometers per hour. It can be observed that, for the OCDM system,  the BER performance of the  MP detector is significantly better than that of the MMSE detector. For example, when the BER equals to $10^{-4}$, OCDM-MP outperforms OCDM-MMSE by an Eb/N0 gain of 3.7dB. Furthermore, the OCDM system, using either MP or MMSE detector, outperforms OFDM-MMSE with much steeper BER-versus-Eb/N0 curves. This is because OCDM can achieve the full diversity of the channel, while the diversity gain that can be obtained by OFDM is almost negligible when the normalized Doppler-shift is much less than 1.

	\begin{table}[htbp] \normalsize 
		\centering
		\caption{ System parameters of the terrestrial radio OCDM} 
		\begin{tabular}{l|l} 
			\hline
			Parameter & value \\
			\hline
			Carrier frequency & 5 GHz\\
			Bandwidth & 15.360 MHz \\
			No. of chirps (or subcarriers) ($N$) & 256 \\
			Constellation & BPSK, 4-QAM  \\
			Symbol duration & 19.27 $\mu$S \\
			Guard interval & 2.6 $\mu$S \\
			UE Speed & 300, 500 Kmph \\
			\hline
		\end{tabular} 
		\label{table-simParas}
	\end{table}
	
	\begin{table}[htbp] \normalsize 
		\centering
		\caption{ Delay Power profile of the Extended Vehicular A Model} 
		\begin{tabular}{c|c} 
			\hline
			Path delay (nS)& Relative power (dB)\\
			\hline
			0 & 0 \\
			30 & -1.5 \\
			150 & -1.4\\
			310 &  -3.6\\
			370 &  -0.6\\
			710 & -9.1\\
			1090 & -7.0\\
			1730 & -12.0\\
			2510 & -16.9\\
			\hline
		\end{tabular} 
		\label{table-radio-DelayPower}
	\end{table}

	It should be noted that, since the speed of radio waves is $10^6\sim10^7$ times that of mobile UE, the product of the delay spread $S_t$ and  Doppler spread $S_f$  of the radio channel is often much less than 1 (i.e., the coherence time is much larger than the delay spread). This kind of wireless channels are usually called \emph{under-spread} LTV channels. In the above simulation, the product $S_tS_f$ of the channel is only 0.01. Hence, it is a typical \emph{under-spread} LTV channel. Compared to \emph{under-spread} channels, \emph{over-spread} channels are much more challenging. For an  \emph{over-spread}  channel, owing to its severe double spreads both in time domain and in frequency domain, the product of $S_tS_f$ is generally greater than 1. In the following simulations, we will evaluate the effectiveness of the proposed OCDM receiver under \emph{over-spread} channels.
	
	Here, we consider the scenario of mobile wireless communication under UWA channnels, where the transmit transducers or receive hydrophones are mounted on moving platforms (e.g. ships or unmanned underwater vehicles). The carrier frequency, the signal bandwidth, the number of chirps used in OCDM or sub-carriers in OFDM, the symbol constellation for data bits mapping, the duration of symbol with CP,  the speed of the relative motion between the transmitter and receiver, the speed of sound in water,  the guard interval,  the speed of the relative motion between the transmitter and the receiver, and the speed of sound in water are listed in Table \ref{table-simParas-UWA}. The delay-power profile of the simulation channel is provided in Table \ref{table-UWA-DelayPower}. Corresponding to the relative speed of the transmitter and receiver, the channel's maximum Doppler shift $v_{\max}$ is 177.8Hz, which is 7.1 after normalized by $1/T$.  The Doppler shifts of the channel paths are IID random numbers. The Doppler shift of the $i$-th path is  $v_i$ = $v_{\max} \cos {\theta}_i$,  where ${\theta}_i$ is uniformly distributed on the interval $\left[ -\pi /2, \pi /2 \right]$. This is an overspread channel, since the product of the Doppler spread and the delay spread is 5.22, much greater than 1.
	
	\begin{table}[htbp] \normalsize 
		\centering
		\caption{ System parameters of the UWA OCDM} 
		\begin{tabular}{l|l} 
			\hline
			Parameter & value \\
			\hline
			Carrier frequency & 24 KHz\\
			Bandwidth & 3.2 KHz \\
			No. of chirps (or subcarriers) ($N$) & 128 \\
			Constellation & BPSK, 4-QAM  \\
			Symbol duration & 55 mS \\
			Guard interval & 15 mS \\
			Transceiver moving speed & 40 kmph \\
			Speed of sound in water &  1500 mph\\
			\hline
		\end{tabular} 
		\label{table-simParas-UWA}
	\end{table}
	
	\begin{table}[htbp] \normalsize 
		\centering
		\caption{ Delay Power profile of the UWA channel in simulation} 
		\begin{tabular}{c|c} 
			\hline
			Path delay (mS)& Relative power (dB)\\
			\hline
			0 & 0 \\
			0.6 & -0.6 \\
			1.3 & -1\\
			2.2 &  -1.3\\
			6.9 &  -2.8\\
			7.5 & -4.2\\
			8.1 & -3.5\\
			13.1 & -6.2\\
			13.8 & -7.3\\
			14.7 & -8.1 \\
			\hline
		\end{tabular} 
		\label{table-UWA-DelayPower}
	\end{table}

	\begin{figure}[!t]
		\centering
		\includegraphics[width=8.8cm]{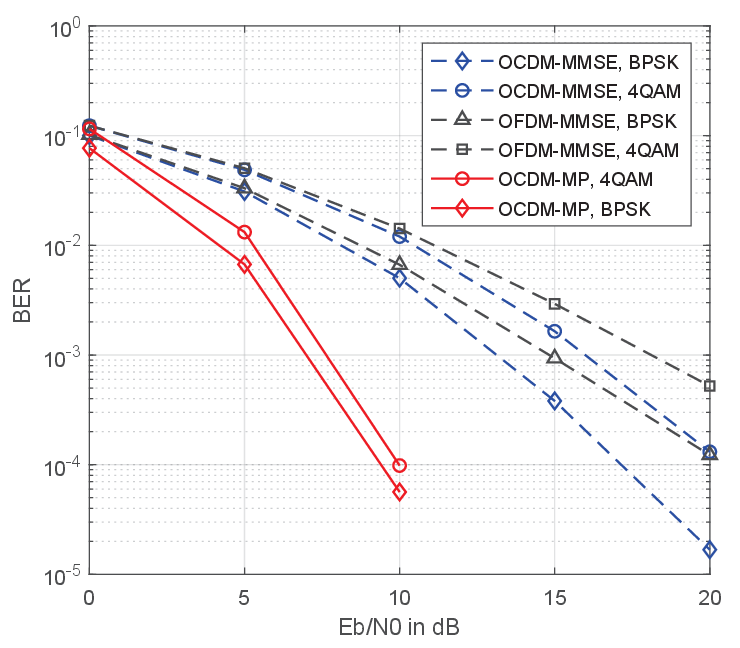}
		\caption{BER performance comparison between OCDM-MP, OCDM-MMSE, and OFDM-MMSE under UWA channel model with transceiver speed 40 kmph.}\label{fig:UWA_BER}	
	\end{figure}
	
	Fig. \ref{fig:UWA_BER} shows the BER performance comparison between OCDM-MP, OCDM-MMSE, and OFDM-MMSE under the above UWA channels model. Considering the fractional Doppler shift $\kappa_i$ can be any number on the interval  $\left(-0.5, 0.5 \right]$, we set  $M_i=10$ in formula  (\ref{eqn:fracDoppler-006}) for all the physical paths in the simulation. The damping factor $\Delta$ and the allowed number of iterations of the MP algorithm are set to be 0.6 and 20 respectively. It can also be observed from the figure that the BER of OCDM-MP is significantly lower than that of OCDM-MMSE and OFDM-MMSE. It should be noted that, since the path normalized Doppler can be greater than 1 in UWA channel, in the OFDM system, a symbol modulated on one subcarrier can be delivered to another subcarrier by a physical path. Thanks to the diversity of path attenuation and Doppler shift, OFDM systems can obtain diversity gains in the above UWA channels. This is why as SNR increases, the BER of OFDM-MMSE in Fig. \ref{fig:UWA_BER} decreases faster than that in Fig. \ref{fig:radio_BER}. It also can be observed from Fig. \ref{fig:UWA_BER} that, the BER performance of OCDM-MMSE is still better than that of OFDM-MMSE. This is because both path delays and Doppler  shifts contribute to the diversity gain of OCDM, while the diversity gain of OFDM only comes from Doppler shifts.
	
	\begin{figure}[!t]
		\centering
		\includegraphics[width=8.8cm]{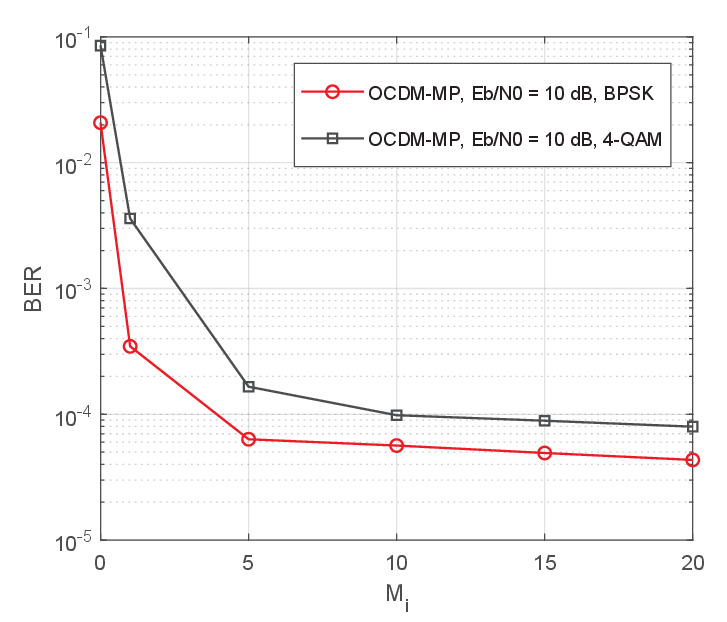}
		\caption{BER performance of OCDM-MP under different $M_i$ when the Eb/N0, the damping factor, and the moving speed of the UWA transceivers is  10dB, 0.6, and  40 kmph respectively.} \label{fig:BER_Mi}	
	\end{figure}
	
	Through formula (\ref{eqn:fracDoppler-006}), the $i$-th physical path with fractional Doppler shift is approximated by $2M_i+1$ virtual paths with integer Doppler shifts. Obviously, larger $M_i$ leads to less approximation error but more computational complexity. And the approximation error is related to the system BER of OCDM-MP. Under the same simulation parameters for Fig. \ref{fig:UWA_BER},  Fig. \ref{fig:BER_Mi} depicts the BER performance of OCDM-MP under different $M_i$ when Eb/N0 = 10 dB. As can be seen from the figure, increasing $M_i$ has a considerable effect on BER when $M_i\leqslant5$,  while the performance improvement brought by increasing $M_i$ is not obvious when $M>10$. Therefore, in this UWA OCDM system, we choose $M_i = 10$ as a good tradeoff between complexity and performance for the message passing based receiver. 
	
\begin{figure}[!t]
		\centering
		\includegraphics[width=8.8cm]{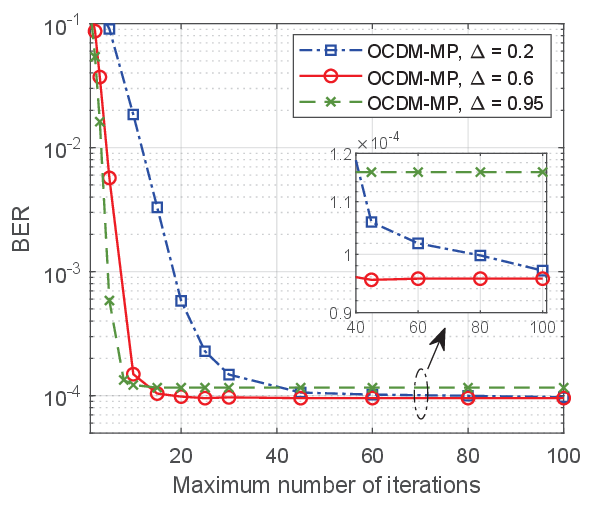}
		\caption{BER versus allowed number of iterations of OCDM-MP under the UWA channel at Eb/N0 = 10 dB and $M_i = 10$.}\label{fig:BER_IterNum}	
	\end{figure}

	Next, we investigate the effect of the \emph{damping factor} $\Delta$ on the BER performance and the computational complexity of the OCDM-MP. In Fig. \ref{fig:BER_IterNum}, we plot the BER versus the allowed number of iterations of the above UWA OCDM-MP system at Eb/N0 = 10 dB and $M_i = 10$.  As described in Algorithm \ref{alg1},  the \emph{damping factor} $\Delta$ is used to control the updating ratio of the message sent from the variable nodes to the observation nodes. As can be seen from the figure, with the smaller \emph{damping factor}  $\Delta$ , the greater allowed number of iterations is required by the MP detector to obtain the best possible BER performance. For instance, when $\Delta = 0.2$, the allowed number of iterations should be set to larger than 40, while a number fewer than 20 is needed for $\Delta = 0.6$. It should be noted that, if $\Delta$ takes a value close to 1, such as 0.95 in the simulation, the achievable BER performance of OCDM-CP would be worse than that using a moderate \emph{damping factor}. Therefore, the choice of $\Delta$ should be a compromise between system complexity and BER performance. 
	
\begin{figure}[!t]
		\centering
		\includegraphics[width=8.8cm]{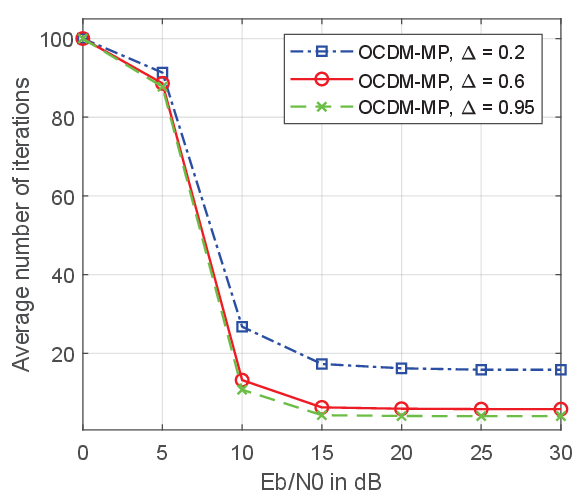}
		\caption{Average number of iterations versus Eb/N0  of the OCDM-MP under the UWA channel with  4-QAM modulation and $M_i =  10$.}\label{fig:NoIter_SNR}	
	\end{figure}

	Fig. \ref{fig:NoIter_SNR} plots the average number of iterations versus Eb/N0 of the OCDM-MP  with 4-QAM modulation, $M_i = 10$, and  the allowed number of iterations set to 100, under the above UWA channel. From the figure, we notice that, no matter how much the \emph{damping factor} is,  at a very low EB/N0, such as 0 dB, the average number of iterations almost reaches the allowed number of iterations, because the algorithm is difficult to converge at low signal-to-noise ratio. At a moderate to high Eb/N0, such as greater than 15 dB,  the average number of iterations does not decrease with the increase of Eb/N0, but related to the \emph{damping factor}. Consistent with Fig. \ref{fig:BER_IterNum}, a greater $\Delta$ will result in fewer iterations (lower  computational complexity). It should be noted that, by changing   $\Delta$  from 0.6 to 0.9, the complexity of the algorithm is negligible, but its BER performance is worse (see in Fig. \ref{fig:BER_IterNum}).  Therefore, in the simulation we choose $\Delta = 0.6$ as an appropriate tradeoff between complexity and BER performance.

	\section{Conclusion}
	
	In this paper,  a message passing based detector was proposed for data symbol iterative estimation over MLMD channels, which are doubly selective in time and frequency domain.  In order to obtain the interference model of the data symbols, we first derived the closed-form expression of the channel matrix in Fresnel domain, using the path attenuation factors, delays, and Doppler shifts of the equivalent base-band channel. 
	Then we proposed a low-complexity computational method for computing an sparse approximation of the Fresnel channel matrix. The approximated Fresnel-domain channel matrix has only $L$ non-zero elements at each row or column. It can be inferred from the expression of the approximated Fresnel matrix that 1) a path delay of   $l_i$ causes the transmit Fresnel-domain symbols to be cyclically shifted by $l_i$ elements; 2) a path Doppler shift of integer $k_i$ causes the transmit data symbols to be cyclically shifted by $k_i$ elements; 3) a fractional Doppler shift $k_i+\kappa_i$ can be approximately equivalent to multiple integer Doppler shifts around $k_i$. Using the approximated channel, we described the Fresnel-domain  input-output relation of the system by a factor graph, based on which the MP detector of OCDM was presented. Finally, we assessed the BER performance of the proposed receiver on two MLMD channel models. The first one is for terrestrial vehicular communication with UE speed of 300 and 500 kmph. The second one is for UWA communication with UUV speed of 1.5 kmph. The simulation results show that the proposed MP receiver outperforms the conventional MMSE frequency-domain equalization based receiver in terms of BER.

	\appendices
	\section{Proof of Lemma \ref{lemma1}}
	Using the  definitions of $\mathbf{\Delta }$ and $\mathbf{\Pi }$ in (\ref{eqn:def_Delta}) and (\ref{eqn:def_PI}), we can write their product $\mathbf{\Delta \Pi }$ as 
	\begin{align}
		\mathbf{\Delta \Pi } &= 
		\left[ \begin{matrix}
			e^{j\frac{2\pi}{N}0}&		&		&		\\
			&		e^{j\frac{2\pi}{N}1}&		&		\\
			&		&		\ddots&		\\
			&		&		&		e^{j\frac{2\pi}{N}\left( N-1 \right)}\\
		\end{matrix} \right] \left[ \begin{matrix}
			0&		\cdots&		0&		1\\
			1&		\cdots&		0&		0\\
			\vdots&		\ddots&		\vdots&		\vdots\\
			0&		\cdots&		1&		0\\
		\end{matrix} \right]  \nonumber \\
		&=  \left[ \begin{matrix}
			0&		\cdots&		0&		e^{j\frac{2\pi}{N}0}\\
			e^{j\frac{2\pi}{N}1}&		\cdots&		0&		0\\
			\vdots&		\ddots&		\vdots&		\vdots\\
			0&		\cdots&		e^{j\frac{2\pi}{N}\left( N-1 \right)}&		0\\
		\end{matrix} \right] \label{eqn:shiftedDelta}. 
	\end{align}
	Each column vector of the matrix in (\ref{eqn:shiftedDelta}) can be regarded as some vector cyclic-shifted by 1 element (realized by multiplying it by the matrix $\mathbf{\Pi}$ ). Then we have
	\begin{align}
		\mathbf{\Delta \Pi } &= 
		\mathbf{\Pi }\left[ \begin{matrix}
			e^{j\frac{2\pi}{N}1}&		\cdots&		0&		0\\
			\vdots&		\ddots&		\vdots&		\vdots\\
			0&		\cdots&		e^{j\frac{2\pi}{N}\left( N-1 \right)}&		0\\
			0&		\cdots&		0&		e^{j\frac{2\pi}{N}N}\\
		\end{matrix} \right]   \nonumber \\
		&= 
		e^{j\frac{2\pi}{N}}\mathbf{\Pi }\left[ \begin{matrix}
			e^{j\frac{2\pi}{N}0}&		\cdots&		0&		0\\
			\vdots&		\ddots&		\vdots&		\vdots\\
			0&		\cdots&		e^{j\frac{2\pi}{N}\left( N-2 \right)}&		0\\
			0&		\cdots&		0&		e^{j\frac{2\pi}{N}\left( N-1 \right)}\\
		\end{matrix} \right] \nonumber \\
		&= e^{j\frac{2\pi}{N}}\mathbf{\Pi \Delta },
	\end{align}
	which completes the proof of Lemma \ref{lemma1}.
	
	\section{Proof of Lemma \ref{lemma2} }
	In this section, we prove the equality of formula (\ref{eqn:lemma2}) in Lemma  \ref{lemma2}, where $k$ is an arbitrary integer. It is obvious that, for $k=0$,  the left-hand side of (\ref{eqn:lemma2}) is $\mathbf{I}_N$, equal to the right-hand side. Hence, we just need to consider the following cases.
	\subsection{The case $k$ being a positive integer}
	Using mathematical induction, we prove that the formula (\ref{eqn:lemma2}) holds when $k$ is a positive integer, by two steps.
	
	1) Prove the equality of (\ref{eqn:lemma2}) when $k=1$, namely, 
	\begin{equation} \label{eqn:proofLemma1-01}
		\mathbf{\Phi \Delta \Phi }^H=e^{j\frac{\pi}{N}}\mathbf{\Pi \Delta }.
	\end{equation}
	At first, let us represent the DFnT matrix $\mathbf{\Phi }$ as a group of column vectors, $\mathbf{\Phi }=\left[ \boldsymbol{\phi }_0,\boldsymbol{\phi }_1,\cdots ,\boldsymbol{\phi }_{N-1} \right]$. Then, the left-hand side of (\ref{eqn:proofLemma1-01}) can be expressed as (\ref{eqn:lemma2Proof-02}) shown on the top of this page.
	\newcounter{mytempeqncnt}
	\begin{figure*}[!t]
		\normalsize
		\begin{align}
			\mathbf{\Phi \Delta \Phi }^H =& \left[ \boldsymbol{\phi }_0,\boldsymbol{\phi }_1,\cdots ,\boldsymbol{\phi }_{N-1} \right] \left[ \begin{matrix}
				e^{j\frac{2\pi}{N}0}&		&		&		\\
				&		e^{j\frac{2\pi}{N}1}&		&		\\
				&		&		\ddots&		\\
				&		&		&		e^{j\frac{2\pi}{N}\left( N-1 \right)}\\
			\end{matrix} \right] \left[ \begin{array}{c}
				\boldsymbol{\phi }_{0}^{H}\\
				\boldsymbol{\phi }_{1}^{H}\\
				\vdots\\
				\boldsymbol{\phi }_{N-1}^{H}\\
			\end{array} \right] \nonumber \\
			=& \sum_{k=0}^{N-1}{e^{j\frac{2\pi}{N}k}}\boldsymbol{\phi }_k\boldsymbol{\phi }_{k}^{H} \label{eqn:lemma2Proof-02}
		\end{align} 
		\hrulefill
		\vspace*{4pt}
	\end{figure*}
	Based on this expression, the $(m,n)$th entry of $\mathbf{\Phi \Delta \Phi }^H$ can be written as
	\begin{equation}
		\left[ \mathbf{\Phi \Delta \Phi }^H \right] _{m,n} = 
		\sum_{k=0}^{N-1}{e^{j\frac{2\pi}{N}k}}\left[ \boldsymbol{\phi }_k \right] _m\left[ \boldsymbol{\phi }_{k}^{H} \right] _n.  \label{eqn:proofLemma1-03}
	\end{equation}
	
	According to the definition of $\mathbf{\Phi }$,  we know that the element $\left[ \boldsymbol{\phi }_n \right] _m$ is equal to $\varphi _{m}^{*}\left[ n \right]$, which is shown in (\ref{eqn_discrete_chirp-2}). Based on this, the formula (\ref{eqn:proofLemma1-03}) can be further written as
	\begin{align} \label{eqn:proofLemma1-04}
		\left[ \mathbf{\Phi \Delta \Phi }^H \right] _{m,n} &= \frac{1}{N}e^{j\frac{\pi}{N}\left( m^2-n^2 \right)}\sum_{k=0}^{N-1}{e^{j\frac{2\pi}{N}k\left[ -m+n+1 \right]}} \nonumber \\
		&= \begin{cases}
			e^{j\frac{\pi}{N}\left( m^2-n^2 \right)}& ,m=\left[ n+1 \right] _N\\
			0&		,\mathrm{others}
		\end{cases},                                                                                                                                                                                          
	\end{align}
	which means that $ \mathbf{\Phi \Delta \Phi }^H $ is a sparse matrix. From (\ref{eqn:proofLemma1-04}), we can rewrite  $ \mathbf{\Phi \Delta \Phi }^H $ as
	\begin{align}
		\mathbf{\Phi \Delta \Phi }^H &= \left[ \begin{matrix}
			0&		\cdots&		0&		e^{-j\frac{\pi}{N}}\\
			e^{j\frac{\pi}{N}1}&		\cdots&		0&		0\\
			\vdots&		\ddots&		\ddots&		\vdots\\
			0&		\cdots&		e^{j\frac{\pi}{N}\left( 2N-3 \right)}&		0\\
		\end{matrix} \right]  \nonumber \\
		&= \mathbf{\Pi }\mathrm{diag}\left\{ \left[ e^{j\frac{\pi}{N}1}, e^{j\frac{\pi}{N}3},\cdots ,e^{j\frac{\pi}{N}\left( 2N-3 \right)},e^{-j\frac{\pi}{N}} \right] \right\} \nonumber \\
		&= e^{j\frac{\pi}{N}}\mathbf{\Pi }\mathrm{diag}\left\{ \left[ e^{j\frac{2\pi}{N}0},e^{j\frac{2\pi}{N}1},\cdots ,e^{j\frac{2\pi}{N}\left( N-1 \right)} \right] \right\} \nonumber \\
		&= e^{j\frac{\pi}{N}}\mathbf{\Pi \Delta },
	\end{align}
	which completes the proof of (\ref{eqn:proofLemma1-01}).

	2) Assuming that the equality of  (\ref{eqn:lemma2}) is true for $k = k_0$, $k_0=1,2,\cdots$, we prove that the equality is also true for $k = k_0 + 1$. Substituting  $k = k_0$ to  (\ref{eqn:lemma2}), we get
	\begin{equation} \label{eqn:assumption-01}
		\mathbf{\Phi \Delta }^{k_0}\mathbf{\Phi}^H=e^{j\frac{\pi}{N}k_{0}^{2}}\mathbf{\Pi }^{k_0}\mathbf{\Delta }^{k_0},
	\end{equation}
	for  $k_0=1, 2, \cdots$.  Then, from (\ref{eqn:proofLemma1-01}) and (\ref{eqn:assumption-01}),  we have
	\begin{align} \label{eqn:lemma2-02}
		\mathbf{\Phi \Delta }^{k_0+1}\mathbf{\Phi }^H  &=  \mathbf{\Phi \Delta }^{k_0}\mathbf{\Phi }^H\mathbf{\Phi \Delta \Phi }^H  \nonumber \\
		&= e^{j\frac{\pi}{N}\left( k_{0}^{2}+1 \right)}\mathbf{\Pi }^{k_0}\mathbf{\Delta }^{k_0}\mathbf{\Pi \Delta }. 
	\end{align}
	Using Lemma 1, the formula (\ref{eqn:lemma2-02}) can be further written as
	\begin{equation}
		\mathbf{\Phi \Delta }^{k_0+1}\mathbf{\Phi }^H=e^{j\frac{\pi}{N}\left( k_0+1 \right) ^2}\mathbf{\Pi }^{k_0+1}\mathbf{\Delta }^{k_0+1}, \label{eqn:lemma2-03}
	\end{equation}
	which completes the proof of step 2). 
	
	According to the results of step 1) and step 2), we can deduce that the equality of formula (\ref{eqn:lemma2}) is always true for any given positive integer $k$. 
	
	\subsection{The case $k$ being a negative integer}
	Similar to the case \emph{A}, given a negative integer $k$,  we prove the correctness of formula (\ref{eqn:lemma2}) by mathematical induction in two steps.
	
	1) Prove the equality of (\ref{eqn:lemma2}) when $k=-1$. Performing matrix inversion on both sides of formula (\ref{eqn:proofLemma1-01}) and using Lemma \ref{lemma1}, we have 
	\begin{align} \label{eqn:proofLemma2-step01}
		\mathbf{\Phi \Delta }^{-1}\mathbf{\Phi }^H&=e^{-j\frac{\pi}{N}}\left( \mathbf{\Pi \Delta } \right) ^{-1} \nonumber \\
		&=e^{-j\frac{\pi}{N}}\left( e^{-j\frac{2\pi}{N}}\mathbf{\Delta \Pi } \right) ^{-1} \nonumber \\
		&=e^{j\frac{\pi}{N}}\mathbf{\Pi }^{-1}\mathbf{\Delta }^{-1},
	\end{align}
	which completes the proof.
	
	2) Assuming that formula (\ref{eqn:lemma2}) holds when $k=k_0$, $k_0= -1, -2, \cdots$, we prove that it also holds when $k = k_0 -1$. Substituting $k=k_0-1$ to the left-hand side of (\ref{eqn:lemma2}), and using the equality of (\ref{eqn:proofLemma2-step01}), we get
	\begin{align} \label{eqn:proofLemma2-04}
		\mathbf{\Phi \Delta }^{k_0-1}\mathbf{\Phi }^H &= \mathbf{\Phi \Delta }^{k_0}\mathbf{\Phi }^H\mathbf{\Phi \Delta }^{-1}\mathbf{\Phi }^H \nonumber \\
		&= \mathbf{\Phi \Delta }^{k_0}\mathbf{\Phi }^H\mathbf{\Phi \Delta }^{-1}\mathbf{\Phi }^H \nonumber \\
		&= e^{j\frac{\pi}{N}k_{0}^{2}+1}\mathbf{\Pi }^{k_0}\mathbf{\Delta }^{k_0}\mathbf{\Pi }^{-1}\mathbf{\Delta }^{-1}.
	\end{align}
	Inverting the matrices on both sides of the equation (\ref{eqn:exchange}), we have
	\begin{equation} \label{eqn:lemma1-inversion}
		\mathbf{\Delta }^{-1}\mathbf{\Pi }^{-1}=e^{j\frac{2\pi}{N}}\mathbf{\Pi }^{-1}\mathbf{\Delta }^{-1},
	\end{equation}
	with which $\mathbf{\Delta }^{k_0}\mathbf{\Pi }^{-1}$ can be written as $e^{j\frac{2\pi}{N}k_0}\mathbf{\Pi }^{-1}\mathbf{\Delta }^{k_0}$. Hence, the formula (\ref{eqn:proofLemma2-04}) can be further simplified as
	\begin{equation}
		\mathbf{\Phi \Delta }^{k_0-1}\mathbf{\Phi }^H = 
		e^{j\frac{\pi}{N}\left( k_0-1 \right) ^2}\mathbf{\Pi }^{k_0-1}\mathbf{\Delta }^{k_0-1},
	\end{equation}
	which completes the proof of step 2).
	
	Therefore, according to the proof in subsections \emph{A} and  \emph{B}, the formula  (\ref{eqn:lemma2}) always holds for arbitrary integer $k$.


\begin{thebibliography}{1}
		\bibitem{DVB}
		H. Sari, G. Karam, and I. Jeanclaude, ``Transmission Techniques for Digital Terrestrial TV Broadcasting,''  \textit{IEEE Communications Magazine}, vol. 33, no. 2, pp. 100-109, 1995.	
		
		\bibitem{WLAN}
		IEEE 802.11 Working Group, ``Wireless LAN Medium Access Control (MAC) and Physical Layer (PHY) Specification,'' 1997.
		
		\bibitem{LTE}
		S. Sesia, I. Toufik, and M. Baker, Eds., \textit{LTE: The UMTS Long Term Evolution}. John Wiley and Sons, 2009.
		
		\bibitem{5G}
		M. Shafi et al., ``5G: A Tutorial Overview of Standards, Trials, Challenges, Deployment, and Practice,'' \textit{IEEE Journal on Selected Areas in Communications}, vol. 35, no. 6, pp. 1201-1221, June 2017.
		
		\bibitem{DS_OFDM_ChnnlEst}
		Z. Tang, R. C. Cannizzaro, G. Leus and P. Banelli, ``Pilot-Assisted Time-Varying Channel Estimation for OFDM Systems,'' \textit{IEEE Transactions on Signal Processing}, vol. 55, no. 5, pp. 2226-2238, May 2007.
		
		\bibitem{DS_OFDM_Eql_Est}
		A. Gorokhov and J. . -P. Linnartz, ``Robust OFDM receivers for dispersive time-varying channels: equalization and channel acquisition,'' in IEEE Transactions on Communications, vol. 52, no. 4, pp. 572-583, April 2004
		
		\bibitem{DS_OFDM_ICI_Eql}
		J. Huang, S. Zhou, J. Huang, C. R. Berger and P. Willett, ``Progressive Inter-Carrier Interference Equalization for OFDM Transmission Over Time-Varying Underwater Acoustic Channels,'' \textit{IEEE Journal of Selected Topics in Signal Processing}, vol. 5, no. 8, pp. 1524-1536, Dec. 2011
		
		\bibitem{DS_MMSE-OFDM} 
		L. Rugini, P. Banelli and G. Leus, ``Simple equalization of time-varying channels for OFDM,'' \textit{IEEE Communications Letters}, vol. 9, no. 7, pp. 619-621, July 2005.
		
		\bibitem{OCDM_ZhaoJian}
		X. Ouyang and J. Zhao, ``Orthogonal Chirp Division Multiplexing,'' \textit{IEEE Transactions on Communications}, vol. 64, no. 9, pp. 3946-3957, Sept. 2016.
		
		\bibitem{OCDM_J_LightWave_2016} 
		X. Ouyang and J. Zhao, ``Orthogonal Chirp Division Multiplexing for Coherent Optical Fiber Communications,'' \textit{Journal of Lightwave Technology}, vol. 34, no. 18, pp. 4376-4386, 15 Sept., 2016.
		
		
		\bibitem{OCDM_SPL_2017} 
		X. Ouyang, O. A. Dobre, Y. L. Guan and J. Zhao, ``Chirp Spread Spectrum Toward the Nyquist Signaling Rate—Orthogonality Condition and Applications,''  \textit{IEEE Signal Processing Letters}, vol. 24, no. 10, pp. 1488-1492, Oct. 2017.
		
		\bibitem{OCDM_FreqD_Pilot_2022_Ma} 
		M. S. Omar and X. Ma, ``Pilot Symbol Aided Channel Estimation for OCDM Transmissions,'' \textit{IEEE Communications Letters}, vol. 26, no. 1, pp. 163-166, Jan. 2022.
		
		\bibitem{OCDM_CFO_2022_Ma} 
		R. Zhang, Y. Wang and X. Ma, ``Channel Estimation for OCDM Transmissions With Carrier Frequency Offset,'' \textit{IEEE Wireless Communications Letters}, vol. 11, no. 3, pp. 483-487, March 2022.
		
		
		\bibitem{OCDM_PAPR_AI_Ma} 
		M. S. Omar, J. Qi and X. Ma, ``Mitigating Clipping Distortion in Multicarrier Transmissions Using Tensor-Train Deep Neural Networks,'' \textit{IEEE Transactions on Wireless Communications}, vol. 22, no. 3, pp. 2127-2138, March 2023.
		
		\bibitem{OCDM_MIMO_ChnnlEst_2023} 
		X. Ouyang, O. A. Dobre, Y. L. Guan and P. Townsend, ``Channel Estimation for Multiple-Input Multiple-Output Orthogonal Chirp-Division Multiplexing Systems,'' \textit{IEEE Transactions on Wireless Communications}, (Early Access), May 2023.
		
		\bibitem{OCDM_UWA_OverResampling_ACCESS2022} 
		P. Zhu, G. Yang, W. Chen, X. Xu and Y. Chen, ``Doppler-Resistant Orthogonal Chirp Division Multiplexing With Multiplex Resampling for Mobile Underwater Acoustic Communication,'' \textit{IEEE Access}, vol. 10, pp. 55151-55163, 2022.
		
		\bibitem{OCDM_UWA_IOT_2023} 
		B. Wang, Y. Wang, Y. Li and X. Guan, ``Underwater Acoustic Communications Based on OCDM for Internet of Underwater Things,'' \textit{IEEE Internet of Things Journal}, vol. 10, no. 24, pp. 22128-22142, 15 Dec.15, 2023.
		
		
		\bibitem{OCDM_UWA_shallowWater_ACCESS2022} 
		B. Yiqi and H. Chuanlin, ``Analysis of Doppler and Multipath on Orthogonal Chirp Division Multiplexing in Shallow Water Acoustic Channel,'' \textit{IEEE Access}, vol. 10, pp. 95928-95935, 2022.
		
		\bibitem{DS_UWA_ChnnlEst_CL_2021} 
		B. Wang and X. Guan, ``Channel Estimation for Underwater Acoustic Communications Based on Orthogonal Chirp Division Multiplexing,'' \textit{IEEE Signal Processing Letters}, vol. 28, pp. 1883-1887, 2021.
		
		
		\bibitem{OCDM_UWA_ACCESS_2020} 
		P. Zhu, X. Xu, X. Tu, Y. Chen and Y. Tao, ``Anti-Multipath Orthogonal Chirp Division Multiplexing for Underwater Acoustic Communication,'' \textit{IEEE Access}, vol. 8, pp. 13305-13314, 2020
		
		\bibitem{OCDM_Radar_Comp_Trans2022}
		L. Giroto de Oliveira, B. Nuss, M. B. Alabd, A. Diewald, M. Pauli and T. Zwick, ``Joint Radar-Communication Systems: Modulation Schemes and System Design,'' \textit{IEEE Transactions on Microwave Theory and Techniques}, vol. 70, no. 3, pp. 1521-1551, March 2022.
		
		\bibitem{OCDM_Radar_ChnnlEst_2023} 
		L. Giroto de Oliveira et al., ``Discrete-Fresnel Domain Channel Estimation in OCDM-Based Radar Systems,'' \textit{IEEE Transactions on Microwave Theory and Techniques}, vol. 71, no. 5, pp. 2258-2275, May 2023.
		
		\bibitem{OCDM_Radar_SPL_2022} 
		Y. Wang, Z. Shi, X. Ma and L. Liu, ``A Joint Sonar-Communication System Based on Multicarrier Waveforms,'' \textit{IEEE Signal Processing Letters} , vol. 29, pp. 777-781, 2022.
		
		\bibitem{OCDM_Analysis_Xiaoli_Ma} 
		M. S. Omar and X. Ma, ``Performance Analysis of OCDM for Wireless Communications,'' \textit{IEEE Transactions on Wireless Communications}, vol. 20, no. 7, pp. 4032-4043, July 2021.
		
		\bibitem{OCDM_MMSE_CPFree_CL_2020} 
		R. Bomfin, M. Chafii and G. Fettweis, ``A Novel Iterative Receiver Design for CP-Free Transmission Under Frequency-Selective Channels,'' \textit{IEEE Communications Letters}, vol. 24, no. 3, pp. 525-529, March 2020.
		
		\bibitem{OCDM_optimal_waveform_CL_2018} 
		R. Bomfin, D. Zhang, M. Matthé and G. Fettweis, ``A Theoretical Framework for Optimizing Multicarrier Systems Under Time and/or Frequency-Selective Channels,'' \textit{IEEE Communications Letters}, vol. 22, no. 11, pp. 2394-2397, Nov. 2018.
		
		\bibitem{OCDM_MMSE_PIC_Bomfin_2018} 
		R. Bomfin, M. Chafii and G. Fettweis, ``Low-Complexity Iterative Receiver for Orthogonal Chirp Division Multiplexing,'' in \textit{2019 IEEE Wireless Communications and Networking Conference Workshop (WCNCW), Marrakech}, Morocco, 2019, pp. 1-6.
		
		\bibitem{DS_Robust_Iter_DS_TWC2021} 
		R. Bomfin, M. Chafii, A. Nimr and G. Fettweis, ``A Robust Baseband Transceiver Design for Doubly-Dispersive Channels,'' \textit{IEEE Transactions on Wireless Communications}, vol. 20, no. 8, pp. 4781-4796, Aug. 2021.
		
		\bibitem{DS_pilot_WCL_YiyinWang_2023} 
		Y. Wang, R. Zhang, L. Yan and X. Ma, ``Pilot Chirp-Assisted OCDM Communications Over Time-Varying Channels,'' \textit{IEEE Wireless Communications Letters}, vol. 12, no. 9, pp. 1578-1582, Sept. 2023.
		
		
		\bibitem{DS_ChannlEst_DD_Domain} 
		S. E. Zegrar and H. Arslan, ``A Novel Delay-Doppler-Based Channel Estimation Method in Doubly-Dispersive OCDM Systems,'' \textit{IEEE Wireless Communications Letters}, Dec. 2023. (Early Access) 
		
		\bibitem{MLMD_highspeedTrain}
		F. Hasegawa et al., ``High-speed train communications standardization in 3gpp 5g nr,'' \textit{IEEE Commun. Standards Mag.}, vol. 2, no. 1, pp. 44–52, Mar. 2018.
		
		\bibitem{MLMD_Vehicle}
		B. Li, Z. Fei, and Y. Zhang, ``UAV communications for 5G and beyond: Recent advances and future trends,'' \textit{IEEE Internet Things J.}, vol. 6, no. 2, pp. 2241–2263, 2018.
		
		\bibitem{MLMD_UWA} 
		W. Li, B. Lin, R. Guo and Z. Hao, ``OTFS for Underwater Acoustic Communications: Frame Design and Channel Estimation,'' \textit{OCEANS 2023 - MTS/IEEE U.S. Gulf Coast}, Biloxi, MS, USA, 2023, pp. 1-9.
		
		\bibitem{LTVChnnlBook}
		F. Hlawatsch, and G. Matz, \emph{Wireless Communications Over Rapidly Time-Varying Channels. Cambridge}. MA, USA: Elsevier Press, 2011.
		
		\bibitem{TseBook}
		D. N. C. Tse and P. Viswanath, \emph{Fundamentals of Wireless Communication}. Cambridge, U.K.: Cambridge Univ. Press, 2005.
		
		\bibitem{OTFS-HongYi}
		P. Raviteja, K. T. Phan, Y. Hong, E. Viterbo, ``Interference Cancellation and Iterative Detection for Orthogonal Time Frequency Space Modulation,'' \textit{IEEE Trans. on Wireless Communications}, vol. 17, no. 10, pp. 6501-6515,  Oct. 2018. 
		
		\bibitem{EVA-Model}
		\emph{Evolved Universal Terrestrial Radio Access (E-UTRA); Base Station (BS) Radio Transmission and Reception, Version 8.6.0}, document 3GPP TS 36.104, Jul. 2009.
		
	\end{thebibliography}

	
	
\end{document}